\documentclass[twocolumn,floatfix, showpacs]{revtex4}

\usepackage[final]{epsfig}
\usepackage{amsmath}
\begin{document}
 
\title{Biased Showers --- a common conceptual Framework for the Interpretation of High $P_T$ Observables in Heavy-Ion Collisions}
 
\author{Thorsten Renk}
\email{thorsten.i.renk@jyu.fi}
\affiliation{Department of Physics, P.O. Box 35, FI-40014 University of Jyv\"askyl\"a, Finland}
\affiliation{Helsinki Institute of Physics, P.O. Box 64, FI-00014 University of Helsinki, Finland}


\begin{abstract}
After the start of the LHC, a plethora of novel observables for jet tomography in heavy-ion collisions has appeared. Many of these studies initially found unexpectedly apparently unaltered jet properties, such as for instance the  momentum distribution of hadrons in jets parallel to the jet axis. This has sparked (sometimes exotic) theoretical efforts to explain these findings. Subsequent results have then shown evidence for modifications when the data is considered in greater detail. However, it has to be realized that almost all current high $P_T$ observables measure conditional probabilities of events, not probabilities. Thus, the correct starting point for their theoretical understanding is Bayes' formula, and the biases introduced by the conditioning are crucial to understanding the outcome. Once this is introduced properly into the modelling process, the initially unexpected results are seen to find a natural explanation in terms of various biases and puzzles largely disappear. In this work, a conceptual framework to classify the various observables according to the types of bias they are sensitive to is presented and illustrated with a large number of case studies ranging from simple jet finding to 2+1 dihadron triggered correlations. 
\end{abstract}
 
\maketitle

\section{Introduction}

The idea of doing jet tomography in ultrarelativistic heavy-ion (A-A) collisions, i.e. to utilize hard processes taking place along the creation of a soft bulk medium to probe both the geometry and the degrees of freedom of the medium has been proposed many years ago \cite{radiative1, radiative2, radiative3, radiative4, radiative5,radiative6}. At the Brookhaven Relativistic Heavy Ion Collider (RHIC), the observables considered to probe this physics was initially the nuclear suppression factor of single inclusive hadrons $R_{AA}$ \cite{PHENIX_first_RAA,PHENIX_RAA,PHENIX_RAA_phi} and the suppression factor $I_{AA}$ of hard back-to-back dihadron correlations \cite{STAR_dihadron_first,STAR_dihadron,STAR_dihadron_DzT}.

Recent high statistics runs at RHIC as well as the significantly larger kinematic reach of heavy-ion experiments at the CERN Large Hadron Collider (LHC) have led to a large variety  of new high $P_T$ observables, in particular also observables involving jet reconstruction using several different jet definitions, among them dijet imbalance measurements  \cite{ATLAS,CMS}, jet-hadron (jet-h) correlations \cite{STAR-jet-h}, h-jet correlations \cite{h-jet}, jet fragmentation function \cite{jet-FF} and jet shapes \cite{jet-shape} or observales utilizing rare electroweak triggers such as $\gamma$-h correlations \cite{gamma-h-STAR,gamma-h-PHENIX} or $\gamma$-jet correlations \cite{CMS_gamma_jet}.

To add to the complexity, jet definitions vary from calorimetric jets in which no unfolding of background fluctuations is done as used e.g. in \cite{ATLAS} to combined track/tower jets with $P_T$ cuts imposed on constituents and a hard tower trigger condition imposed  as used e.g. in \cite{STAR-jet-h}.

In this situation, it is fairly difficult to assemble a picture of what information the various observables actually carry, to what degree they are mutually consistent and what features of models they constrain. The aim of this paper is to improve on this situation by providing a clear conceptual framework in which similarities and differences between the various observables become transparent.

The key observation for this is that the vast majority of observables (with the exception of nuclear modification factors) are measurements of a conditional probability given a trigger condition. The fundamental reason for this is that both hard and electroweak processes are rare, i.e. if there would be no selection of the subclass of events containing hard processes, the background of soft bulk medium physics would dilute all  signatures of hard probes to the point where they would no longer be observable. However, conditional probabilities are well known to be frequently non-intuitive, and the natural starting point for analyzing them is Bayes' formula, which will be utilized in the following.

\section{Observables and conditional probabilities}

\subsection{General considerations}

In perturbative Quantum Chromodynamics (pQCD), the rate of hard scattering processes can be computed with reasonable accuracy once the momentum transfer in the scattering process exceeds a few GeV. The uncertainty principle allows to estimate the timescale for the hard reaction as $\tau \sim E/Q^2$ where $E$ is the energy scale of the final state partons and $Q \sim O(E)$ the virtuality scale. Inserting typical numbers, one finds that hard processes occur before a soft medium can be formed, which is the reason that the pQCD computation of hard processes can safely be assumed to factorize from any medium physics. This property makes high $P_T$ observables a meaningful tomographic probe.

The highly virtual back-to-back partons subsequently undergo a final state shower evolution in which the virtuality scale decreases from its initial high value to a non-perturbative scale via the branching into additional partons. This process in vacuum is well described by MC formulations such as the PYSHOW algorithm of PYTHIA \cite{PYSHOW}. Once at the non-perturbative scale, the parton shower hadronizes and becomes a collimated spray of hadrons. Jet clustering algorithms such as anti-$k_T$ or SIS-cone as provided e.g. by the FastJet package \cite{FastJet} aim to 'undo' the QCD shower evolution and turn the spray of hadrons again into a 'jet', i.e. an object which is a reasonable proxy for the original parton largely free of the complications of shower evolution and hadronization and sensitive to hard physics only.

Measurements of hard probes in the context of heavy ion collisions aim at answering the question how the medium modifies this evolution, i.e. in what way the properties of the shower are different if it evolves inside a medium. If a jet contains $n$ hadrons, since the position space information can not be resolved, the complete theoretically measurable  information about the jet is contained in the momentum space density $\rho_n(P_1, P_2,... P_n)$ and in the knowledge of hadron identities. However, currently the focus is on measurements of  the single particle distribution $\rho_1(P_1) = \int dP_2\dots dP_n \rho_n(P_1,P_2,\dots P_n)$, usually represented as parallel and perpendicular momentum spectra of particles with respect to the jet axis. In the future measurements may also include intra-jet correlations. These would be given e.g. by two particle correlation $C_2(P_1, P_2)$ and three particle correlations $C_3(P_1,P_2,P_3)$ or expressed in terms of subjet fractions. This information may be represented in different form, for instance the integrated jet shape
\begin{equation}
\Psi_{int}(r,R) = \frac{\sum_i E_i \theta(r-R_i)}{\sum_i E_i \theta(R-R_i)}
\end{equation}

 (the integrated flux of energy as a function of angle $r$ with the jet axis of a jet of radius $R$, normalized to the total jet energy) is computable from the angular distribution of hadrons at given energy $dN/d\phi dE$ (as for instance obtained from a correlation measurement) as\begin{equation}
\Psi_{int}(r,R) = \frac{\int_0^r d\phi dE E \frac{dN}{d\phi dE}}{\int_0^R d\phi dE E \frac{dN}{d\phi dE}}.
\end{equation}
 
It is thus theoretically sufficient to measure one representation of the single particle distribution, different representations contain redundant information. However, in practice a jet shape is always conditional on having found a jet in an event, whereas the angular distribution of hadrons obtained from a triggered correlation measurement is conditional on a different trigger condition, and hence the two representations will in practice not conntain precisely the same information. Moreover, no real measurement can resolve the true particle composition of every jet. 

If we use the notation that $P(A|B)$ stands for the probability of event $A$ occurring given another event $B$, the computation of the probability of observing shower properties $S$ (for instance the probability of measuring a shower hadron between 2 and 3 GeV) given a set of trigger conditions $T$ (for example given that a jet is clustered in an energy between 100 and 150 GeV) is written as $P(S|T,M)$ where $M$ stands for the particular model in which the calculation is carried out. Bayes' formula then allows to compute this as

\begin{equation}
\label{E-1}
P(S|T,M) = \frac{P(T|S,M) P(S|M)}{P(T|M)}
\end{equation}

In words, the probability for observing shower properties $S$ given a trigger $T$ is the product of the probability to fulfill the trigger condition in a shower with property $S$ times the probability to generate a shower $S$, divided by the probability to generate a trigger \emph{independent} if property $S$ is realized or not. Since a rate is obtained by multiplying a probability with a repetition frequency, the whole language trivially generalizes to event rates or particle spectra. 

What is measured is usually the left hand side of the equation, somtimes also the denominator of the right hand side (which corresponds to the rate at which the trigger condition is fulfilled). Eq.~(\ref{E-1}) states then that in a large class of measurements, the medium modification as computable in a model $P(S|M)$ is not be observed directly, but rather is distorted through a \emph{bias factor} $\frac{P(T|S,M)}{P(T|M)}$ which is characterized by the trigger condition $T$. This bias can vary a lot, for instance the requirement to find a 100 GeV calorimetric jet leads to a very different bias than the requirement to find a 20 GeV charged hadron. However as these examples indicate, the formalism applies as well to jet finding followed by an analysis of the fragmentation pattern of the clustered jet \cite{jet-FF} (in which case the jet finding constitutes the trigger condition and the observable is the momentum spectrum of the shower parallel to the jet axis) as to $I_{AA}$ in triggered h-h correlations (in which the requirement to find a hard hadron constitutes the trigger condition and the ratio of parallel momentum spectra of correlated hadrons in medium over vacuum is the observable).

This suggests a clear strategy to make the information content of measurements apparent and comparable: Measure the observable (e.g. the single particle distribution of jet constituents) in the same representation in all measurements and view the different trigger condition as a variation of the bias factor. Tomographic information is then contained in the way the observable responds to a change of the bias factor. 

\subsection{Theoretical formulation of in-medium showers}

As discussed in detail in \cite{Constraining}, modelling of the medium modification of a shower involves a procedure to compute the medium-modified fragmentation function (MMFF). The MMFF can be written in the rather general form $D_{i \rightarrow h} (z, E, Q_0^2 | T_1(\zeta), T_2(\zeta), \dots T_n(\zeta))$, where it describes the distribution of hadrons $h$ given a parton $i$ with initial energy $E$ and initial virtuality $Q_0^2$ where the hadron energy $E_h = z E$ and the parton has traversed a medium along the path $\zeta$ where $T_i(\zeta)$ are the medium transport coefficients relevant for the process. 

Since the MMFF should approach the usual vacuum fragmentation function when the transport coefficients vanish, the properties of a vacuum shower are largely determined by just three parameters --- the shower-initiating parton type $i$, its initial energy $E$ and virtuality $Q_0$. In contrast, the determination of medium modifications in principles require $n$ different functions $T_i(\zeta)$. However, it turns out that in practice three are most relevant: $\hat{q}$ (the medium-induced perpendicular momentum squared per unit pathlength, effectively corresponding to a medium-induced virtuality), $\hat{e}$ (the mean momentum transfer parallel to the parton direction into the medium per unit pathlength, effectively corresponding to parton energy loss) and $\hat{e}_2$ (the variance of the energy loss) \cite{AbhijitReview}. Moreover, in many models it turns out that the full functional dependence of the transport coefficient is not needed but rather the line integral along the parton path $\zeta(\tau)$ as $M_1 = \int d\zeta T_i(\zeta)$ and the line integral along the path with a weight given by the pathlength $\zeta$, i.e. $M_2 = \int d\zeta \zeta T_i(\zeta)$ are to good accuracy sufficient \cite{ASWScaling,YaJEM2}. This implies that the medium modification of a shower can be characterized reasonably well by the set of $M_1(\hat{q}), M_2(\hat{q}), M_1(\hat{e}), M_2(\hat{e}), M_1(\hat{e}_2), M_2(\hat{e}_2)$ which now contain all tomographic information.

Thus, ideally one would like to compare $D_{i \rightarrow h} (z, E, Q_0^2 | M_1(\hat{q}), \dots)$ with a measurement to deduce the tomographic information on the properties of the medium. Photon-triggered correlation come in practice closest to this ideal as they can provide stringent constraints on $E$, but they leave $Q_0^2$ and the location of the initial vertex and hence the set $M_i$ unconstrained.

A number of models for the computation of the MMFF are proposed. Historically, the computation has often been based on the leading parton energy loss approximation in which the virtuality evolution of the shower is not treated explicitly and the focus is only on induced radiation from the leading parton \cite{QuenchingWeights,AMY-1,AMY-2,radiative5,WHDG, radiative6}. Since this approximation is not well suited for the interpretation in terms of conditional probabilities, we will not consider it here. Alternatively, Monte-Carlo (MC) codes for in-medium shower evolution \cite{YaJEM2,YaJEM1,JEWEL,Q-PYTHIA,MARTINI}, parton cascade \cite{VNI} as well as analytical approaches \cite{HT-DGLAP,Vitev} exist.

\subsection{Initial state and final state biases}

In order to directly test a model of jet quenching, it would be desirable if an observable could be constructed in such a way that the vacuum shower model parameters $(i,E,Q_0)$ or the medium parameter moments take fixed values. In this case, the theoretical model would only ever need to consider events which fulfill the trigger condition by construction, rendering the bias factor  $\frac{P(T|S,M)}{P(T|M)}$ identically unity, which simplifies the computation tremendously. This is the reason schematic investigations and toy models follow this strategy. In other words, if one could prepare a situation in which a quark with specified energy propagates through a given length of medium with given density, jet tomography through comparison of experiment and theory would be easy to do.

Unfortunately, experimental measurements are hardly ever conditioning on initial state properties of the shower, in which case the bias factor is different from unity and a model to compute the MMFF is insufficient to compare with data. Instead, experimental trigger conditions usually key on some property of the observed final state after shower evolution and hadronization.

Consider the term $P(T|M)$ which can be written as
\begin{equation}
\label{E-2}
P(T|M) = \sum_{S'} P(T|S',M) P(S'|M)
\end{equation}
using the fact that probabilities normalize to unity. Eq.~(\ref{E-2}) states that in order to compute the rate at which the trigger condition is fulfilled, we need not only compute the shower $S$ exhibiting a particular property we are interested in but in fact all possible shower configurations and medium modifications $S'$ which are allowed by the physics of the collision and do an appropriate sum over them. It is this need to compute all possible initial configurations and check them for the trigger condition in the final state which makes a proper computation vastly more complicated than a toy model estimate.

In practical terms, this means that in order to compute observables which can be compared with experiment, an in-medium shower model needs to be embedded into a framework simulating the hard process and the evolution of the surrounding medium (for a detailed discussion see \cite{Constraining}).

The final state trigger condition than maps (in a model- and embedding-dependent way) into distributions in the space of initial shower parameters. 

\subsection{Monte-Carlo treatment of jet quenching}

\begin{figure}[htb]
\epsfig{file=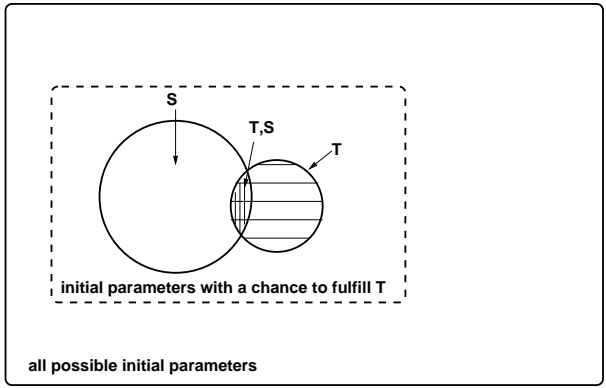, width=8cm}
\caption{\label{F-MC}A schematic illustration of initial parameter space sampling for a conditional probability observable.}
\end{figure}

Let us for illustration consider a MC description of jet quenching. Biases are taken into account by generating events according to the full abailable space of initial parameters with correct weight assigned to the individual contributions, then searching which of these events fulfill the trigger condition in the final state and analyzing only this subset of event to obtain the observable. Pictorially this is shown in Fig.~\ref{F-MC}. Ultimately interesting for the observable are only the two shaded regions, i.e. the class of events which fulfills the trigger T and the class of events which fulfills T and shows property S.

However, the computational problem is that the full range of events generated by sampling all the available initial parameter space is usually so huge that a naive application of the above strategy is bound to be so slow that it is useless in practice. The challenge resulting from this is to introduce an intemediate layer, i.e. to understand the bias structure in such a way that only initial parameter ranges are sampled which have a reasonable chance to lead to a trigger in the final state. Pictorially, this corresponds to drawing the dashed line as closely as possible to the intersecting circles without actually cutting parameter space out of a circle (which would introduce an unphysical sampling bias). In this way, computations become feasible. This illustrates that good knowledge of the mapping of final state conditions to initial state parameters in terms of biases is not only conceptually important, but also has consequences of immediate practical value.

\section{Types of biases}

Following the discussion in \cite{Dihadron2}, we can classify the various biases induced by a trigger condition on the final state of a hard event as follows: First, there are biases on the structure of the hard pQCD event itself which act even in vacuum. These have to do with the relation between hadronic (or jet) and parton kinematics dependent on parton type. Once a medium is present, the correlation of the strength of the medium modification with the density of the medium and the time spent in the medium leads to additional biases on the reaction geometry. Since all these biases act on the hard event itself rather than the final state shower, they affect both trigger side and away side simultaneously. This can be contrasted with shower biases, which affect the structure of the shower evolution itself and do not bias the kinematics or position of the hard event and are thus always only relevant for the trigger side.

In this section, we review qualitatively the effects of the most relevant biases, which we study later with case studies in a full modelling framework. In order to illustrate the isolated effects of the various biases, the examples shown outside the full case studies are theoretical situations in which the initial state of the shower is given, whereas the later experimentally relevant case studies show results given an observed final state.

\subsection{Biases in vacuum showers}

Neither a hadron nor a jet typically contain all the initial parton energy $E$. In the case of a hadron, this is because of the production of subleading hadrons as well as hadron species which are not registered by the detector in the shower. In the case of a jet, the reason is typically the production of hadrons at large angles with the jet axis which correspond to energy flow outside the jet radius $R$, but for instance in charged jets also neutral hadron production in the shower constitute an energy component not part of the jet. 

For both jet and hadron, the relation of observed energy to parton energy can be written into the form $E_{obs} = z_{had/jet} E$. Typically, the chief difference between jet and hadron observation is that a jet tends to recover a higher fraction of the parton energy than a single hard hadron, i.e. $\langle z_{jet} \rangle > \langle z_{had} \rangle$ where the average is done over many showers with a fixed parton energy $E$.

\begin{figure}[htb]
\epsfig{file=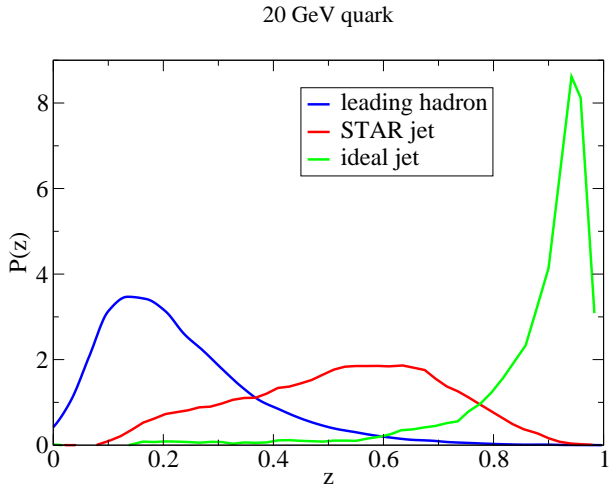, width=8cm}
\caption{\label{F-Pz} (Color online) The probability density $P(z)$ to observe a trigger object with fraction $z = E_{obs}/E$ given an initial parton energy $E$ and an observed trigger energy $E_{obs}$ for various possible trigger objects, shown for the example of a fragmenting 20 GeV quark.}
\end{figure}

This is illustrated in Fig.~\ref{F-Pz} where $P(z)$, the probability to observe the fraction $z$ of the original energy of a 20 GeV quark  in the final state is shown for three different objects: 1) the leading hadron if it is $\pi^+, \pi^-, \pi^0, K^+, K^-, p$ or  $\overline{p}$ 2) a STAR jet definition \cite{STAR-jet-h} where all particles which are $\pi^+, \pi^-, \pi^0, K^+, K^-, p$ or  $\overline{p}$ or $\gamma$ and have $P_T > 2$ GeV are clustered usign the anti-$k_T$ algorithm with a radius of $R=0.4$ and 3) an ideal jet definition where all particles, regardless of PID or $P_T$, are clustered with anti-$k_T$ using $R=0.4$.

It is evident that the leading hadron in this kinematical regime typically carries only about 15\% of the original parton energy whereas on the other end of the spectrum clustering into a jet ideally recovers typically 95\% of the energy. Jet definitions matching realistic experimental conditions fall between the two cases.

A {\itshape kinematic bias} arises then because in an experimental context $P(z)$ is typically not probed for fixed parton energy, but rather folded with the steeply falling primary parton production spectrum which can be computed in pQCD and typically falls approximately like a power $1/p_T^n$ with $n=7..8$ at RHIC kinematics and $n=4..5$ at the LHC.  A trigger energy requirement then demands a fixed $E_{obs} = zE$ where both $z$ and $E$ are allowed to vary event by event. For the ideal jet described above where $P(z) \approx \delta(z-1)$, the bias is negligible and $E_{obs}$ approximately corresponds to the parton energy. For a hadron trigger however, both $E$ and $z$ prefer to be individually small, yet their product is forced to a certain value. As a result, $E_{obs}$ maps to a characteristic range in $E$ which depends on $n$ and the details of $P(z)$, i.e. the distribution of parton energies contributing to a trigger is no longer the primary pQCD spectrum but becomes biased. In \cite{TriggerBias} this is referred to as 'trigger bias', however in the following we will use this term in a more general sense referring to any bias introduced by a trigger condition in either vacuum or medium.

Another part of the kinematic bias is related to the fact that due to higher order pQCD effects and nuclear initial state effects a hard parton pair is never exactly back to back. These effects can be approximated by introducing a randomly oriented vector ${\bf k_t}$ with a Gaussian distribution in magnitude which is added to the pair momenta. A trigger condition then biases this {\itshape a priori} randomly oriented vector to be pointing towards the trigger direction \cite{Dihadron2}. 

The {\itshape parton type bias} then has to do with the fact that the functional form of $P(z)$ depends on the shower-initiating parton type: On average, quarks fragment into harder and more collimated showers than gluons. As a result, any trigger condition corresponding to an observed energy is more likely to be fulfilled by a quark than by a gluon. Thus, on the trigger side the fraction of quark jets is generically enhanced as compared to an unbiased pQCD spectrum. How the bias acts on the away side depends on the kinematical situation. In a regime where the subprocess $qg \rightarrow qg$ is dominant, enhancing the near side quark fraction biases the away side towards gluon jets \cite{Dihadron2} which is relevant for instance for the 5-20 GeV momentum regime at RHIC. 

The biases in vacuum are summarized in Tab.~\ref{T-BiasesV}.

\begin{table}[htb]
\begin{tabular}{|l|l|}
\hline
bias & cause\\
\hline
\hline
kinematic & the relationship between parton and trigger\\ 
& energy results from both spectrum and\\
& fragmentation process\\ 
parton type & gluon jets are softer and less likely to\\
& fulfill a trigger condition\\
\hline
\end{tabular}
\caption{\label{T-BiasesV}The various biases in vacuum}
\end{table}

\subsection{Biases in the medium}

Medium modifications to the shower structure generically tend to equilibrate the shower, i.e. they drive the kinematical properties of shower partons closer to those of medium partons. This implies that medium-modified showers are softer and broader, i.e. more weight in $P(z)$ shifts to lower $z$. As a result, the kinematical bias is changed in a medium, the same $E_{obs}$ maps on average to a higher $E$ in a medium than in a vacuum.

The strength of the medium modification is (up to coherence effects which are important in detail) driven by the number of interactions with the medium, which is a function of the medium density, the coupling strength of shower partons to the medium and the time/length of the shower spent in medium. Out of these, the coupling strength is relevant for a modification of the parton type bias by the medium: Since gluons interact with a factor of 9/4 more strongly with color charges, the medium modification of gluon jets is correspondingly stronger than that for quark jets. Note that the factor 9/4 does not accurately describe the difference between quark and gluon parton showers, as for instance a gluon may split into a $q\overline{q}$ pair which after decoherence interacts as independent quark color charges. However, the dominant radiation pattern in a shower, both for quarks and gluons, is the emission of soft gluons which preserves the identity of the leading parton, and thus gluon jets in practice have a stronger interaction with the medium than quark jets, although the real difference is somewhat smaller than 9/4.
For this reason, triggered objects are even more biased to be quark jets than this is already the case in vacuum. This effect is sometimes referred to as \emph{gluon filtering}.

The combined effect of medium density and pathlength of a parton through the medium leads to a \emph{geometrical bias} on the position of the vertex leading to the triggered event in the transverse plane in position space. Vertices leading to triggered events have a tendency to be close to the medium surface, with the trigger parton travelling outward. This implies that the same effect biases the away side parton to have a longer than average pathlength in the medium.

\subsection{Shower biases}

While all biases discussed so far affect properties of the hard event itself, and thus refer equally to near and away side, there are also biases which affect the trigger parton side only. Those are here referred to as \emph{shower biases}. For instance, requiring that a single hard hadron is produced in a shower restricts the phase space for associated hadron production via the conservation of energy and momentum. Generically, shower biases make observables more robust against medium modifications, as a shower bias implies that there are properties of the shower which are by the trigger condition protected against medium modifications.

A list of the medium-induced biases discussed in this work is given in Tab.~\ref{T-Biases}.

\begin{table}[htb]
\begin{tabular}{|l|l|}
\hline
bias & cause\\
\hline
\hline
kinematic & medium-induced radiation changes relation \\
& between parton and trigger energy\\
\hline
parton type & medium interaction preferentially\\
& suppresses gluon jets\\
\hline
geometry & short in-medium pathlengths are more\\
& likely to fulfill trigger condition\\
\hline
shower & strongly broadened and softened showers\\
& are unlikely to lead to a trigger\\
\hline
\end{tabular}
\caption{\label{T-Biases}The various medium-induced biases}
\end{table}

\section{Model description}

In order to illustrate the qualitative remarks made above quantitatively, we will in the following show results obtained with the in-medium shower evolution code YaJEM \cite{YaJEM2,YaJEM1} in its latest version YaJEM-DE \cite{YaJEM-DE} which gives a fair account of a large number of observables both at RHIC and at LHC \cite{Constraining}. 

YaJEM is a tool to obtain the MMFF given initial parton energy and a path through the medium, hence for a complete model description of a hard process in a medium also the medium evolution and the pQCD process  have to be taken into account.

\subsection{The perturbative hard process}

Any simulation of hard events inside a heavy-ion collision which is not a theoretical quantity with a fixed initial state but refers to an experimentally observed final state must start with the computation of the probability to obtain certain parton momenta and types from the hard process itself.

In LO pQCD, the production of two hard partons $k,l$ 
is described by
\begin{equation}
\label{E-2Parton}
  \frac{d\sigma^{AB\rightarrow kl +X}}{dp_T^2 dy_1 dy_2} \negthickspace 
  = \sum_{ij} x_1 f_{i/A}(x_1, Q^2) x_2 f_{j/B} (x_2,Q^2) 
    \frac{d\hat{\sigma}^{ij\rightarrow kl}}{d\hat{t}}
\end{equation}
where $A$ and $B$ stand for the colliding objects (protons or nuclei) and 
$y_{1(2)}$ is the rapidity of parton $k(l)$. The distribution function of 
a parton type $i$ in $A$ at a momentum fraction $x_1$ and a factorization 
scale $Q \sim p_T$ is $f_{i/A}(x_1, Q^2)$. The distribution functions are 
different for free protons \cite{CTEQ1,CTEQ2} and nucleons in nuclei 
\cite{NPDF,EKS98,EPS09}. The fractional momenta of the colliding partons $i$, 
$j$ are given by $ x_{1,2} = \frac{p_T}{\sqrt{s}} \left(\exp[\pm y_1] 
+ \exp[\pm y_2] \right)$.
Expressions for the pQCD subprocesses $\frac{d\hat{\sigma}^{ij\rightarrow kl}}{d\hat{t}}(\hat{s}, 
\hat{t},\hat{u})$ as a function of the parton Mandelstam variables $\hat{s}, \hat{t}$ and $\hat{u}$ 
can be found e.g. in \cite{pQCD-Xsec}.

To account for various effects, including higher order pQCD radiation, transverse motion of partons in the nucleon (nuclear) wave function and effectively also the fact that hadronization is not a collinear process, the distribution is commonly folded with an intrinsic transverse momentum $k_T$ with a Gaussian distribution, thus creating a momentum imbalance between the two partons as ${\bf p_{T_1}} + {\bf p_{T_2}} = {\bf k_T}$.

In a MC description of the process, Eq.~(\ref{E-2Parton}) is sampled to generate the parton type and momentum of the back-to-back pair. Subsequently the intrinsic ${\bf k_T}$ imbalance is sampled and added to the parton pair momentum. In correlation studies, one of the partons is randomly picked as a trigger candidate.

\subsection{Medium-modified fragmentation}

Hard vertices are assumed to be distributed with a binary overlap profile as appropriate for LO pQCD parton production, i.e. the {\itshape a priori} probability density for finding a vertex in the transverse $(x,y)$ plane is given 

\begin{equation}
\label{E-Profile}
P(x_0,y_0) = \frac{T_{A}({\bf r_0 + b/2}) T_A(\bf r_0 - b/2)}{T_{AA}({\bf b})},
\end{equation}
where the thickness function is given in terms of Woods-Saxon distributions of the the nuclear density
$\rho_{A}({\bf r},z)$ as $T_{A}({\bf r})=\int dz \rho_{A}({\bf r},z)$ and $T_{AA}({\bf b})$ is the standard nuclear overlap function $T_{AA}({\bf b}) = \int d^2 {\bf s}\, T_A({\bf s}) T_A({\bf s}-{\bf b})$ for impact parameter ${\bf b}$. 

In the MC procedure, we place the parton pair at a probabilistically sampled vertex $(x_0,y_0)$ sampled from this distribution with a random orientation $\phi$ with respect to the reaction plane. We rotate the event for the purpose of extracting vertex distributions such that the vector of the trigger candidate parton defines the $-x$ direction. In studies of explict dependence of observables on the angle of the parton with the various $v_n$ event planes we would use the relevant event plane angle instead and would consider only parton propagation with a set angle to the event plane.

The event is now embedded into a hydrodynamical description of the medium (\cite{hydro2d} for the RHIC case and the extrapolation of this scenario to larger $\sqrt{s}$ in the LHC case \cite{RAA-LHC}) which allows to extract e.g. the energy density $\epsilon(\zeta)$ at any point of the propagating parton path $\zeta$. 

In the absence of a medium, YaJEM is identical to the PYSHOW algorithm \cite{PYSHOW} which evolves partons as a series of $a\rightarrow bc$ branchings in the energy fraction $z = E_b/E_a$ and the virtuality $t = \ln(Q_a^2)/\Lambda_{QCD}^2$ with $\Lambda_{QCD} = O(300)$ MeV. In YaJEM, it is assumed that the virtuality $Q_a^2$ and energy $E_a$ of any intermediate shower parton $a$ is modified by the medium via two transport coeffients, $\hat{q}$ and $\hat{e}$ as

\begin{equation}
\label{E-Qgain}
\Delta Q_a^2 = \int_{\tau_a^0}^{\tau_a^0 + \tau_a} d\zeta \hat{q}(\zeta)
\end{equation}

and

\begin{equation}
\label{E-Drag}
\Delta E_a = \int_{\tau_a^0}^{\tau_a^0 + \tau_a} d\zeta \hat{e}(\zeta).
\end{equation}

To evaluate these equations requires a mapping of the shower evolution of PYSHOW in momentum space to the hydrodynamical evolution in position space and a model of the transport coefficients as a function of thermodynamical properties of the medium. 

The temporal structure of the shower evolution can be parametrically recovered by uncertainty arguments. The mean lifetime of a virtual parton $b$  coming from a parent $a$ is hence given as 

\begin{equation}
\label{E-Lifetime}
\langle \tau_b \rangle = \frac{E_b}{Q_b^2} - \frac{E_b}{Q_a^2}.
\end{equation} 

In the MC simulation of the shower, the actual lifetime is determined from this mean value according to the probability distribution

\begin{equation}
\label{E-RLifetime}
P(\tau_b) = \exp\left[- \frac{\tau_b}{\langle \tau_b \rangle}  \right].
\end{equation}

For the relation between transport coefficients and hydrodynamical parameters, 

\begin{equation}
\label{E-qhat}
\hat{q}[\hat{e}](\zeta) = K_Q[K_E] \cdot 2 \cdot [\epsilon(\zeta)]^{3/4} (\cosh \rho(\zeta) - \sinh \rho(\zeta) \cos\psi)
\end{equation}

is assumed where $\rho$ is the transverse flow rapidity of the medium, $\psi$ the angle between parton direction and medium flow direction and $K_Q$ and $K_E$ are two free parameters parametrizing the strength of the coupling of medium and shower partons. In this expression, $\epsilon^{3/4}$ represents a quantity with the dimensions of $\hat{q}$ and in an ideal gas parametrically corresponds to the medium density, whereas the latter factor accounts for the Lorentz contraction (and hence effective density increase) of the volume passed by the hard parton.

Following the procedure in \cite{YaJEM-DE}, these are adjusted to $K_Q = 0.8 K$ and $K_E = 0.1 K$ (corresponding to about a 10\% elastic energy loss contribution leading to direct energy transfer into the medium) and $K$ is fit to the nuclear suppression factor $R_{AA}$ in 0-10\% central Au-Au collisions at RHIC. 

Note that in this work, any results presented are intended as illustration of qualitative effects and the order of magnitude of various biases. Hence no attempt is made to obtain a good fit to any other data set by either fitting $K$ to a more extended set of data or by exploiting the freedom to choose a different fluid dynamical description of the medium. For this reason comparison with data (where it is available) is left for future work.

Changing the kinematics of evolving shower partons according to Eqs.~(\ref{E-Qgain},\ref{E-Drag}) in YaJEM results in a medium-modified parton shower. The resulting distribution of partons is then passed to the Lund model \cite{Lund} to compute the non-perturbative hadronization.

\subsection{Analysis}

The resulting output in terms of medium-modified hadron showers is now analyzed if the trigger condition is fulfilled. Since the event record at this point contains the full information on hadron PID and momenta, in principle any set of cuts can be evaluated (in practice the statistics may become too low).

In case the trigger is a hadron, the test for the trigger condition is trivial. In the case of a jet trigger, the resulting event record is clustered with the anti-$k_T$ algorithm of the FastJet package \cite{FastJet}. At this point particles computed from a bulk hydrodynamical event could be inserted and clustered together with the hard event to study the influence of background fluctuations. This however is computationally very expensive and not done here --- throughout this work, it is assumed that any background fluctuations are sufficiently trivial to be removed. Dependent on the trigger conditions, this may not be a good assumption in practice (see e.g. \cite{BgFluct,LeticiaBG} for studies of the influence of the soft background on observables).

In the case a jet trigger in combination with conditions on PID or constituent momenta is required, all particles not fulfilling the conditions are removed before clustering is done.

\subsection{Relevance of the results}

Despite the fact that the following case studies are performed with a specific parton-medium interaction model, YaJEM-DE, and for a specific choice of medium evolution, the qualitative conclusions drawn about the role of biases in hard observables require a significantly less stringent set of assumptions.

The medium-induced biases will appear acting in the same direction as illustrated by YaJEM-DE for any model which has the following characteristics: 1) the medium on average softens fragmentation in a shower 2) the medium on average broadens the perpendicular distribution of hadrons in a shower 3) the effects of softening and broadening increase monotonously with medium density and in-medium pathlength 4) gluons couple more strongly to the medium than quarks.

Most current models of parton-medium interactions exhibit these traits, so the following qualitative statements can be expected to hold fairly generically. However, quantitatively the relative strength of different biases depends on specific model assumptions.

\section{Shower biases}

A trigger condition may refer to one or both showers generated in a back-to-back hard event. If the shower on the trigger side is studied, in general the trigger condition biases the shower evolution itself. The shower bias is however absent when the trigger condition is evaluated for one parton (the 'near side') whereas the measurement of jet properties is done for the other parton (the 'away side'). Typical examples for measurements in which shower biases occur are the near side associated hadron distribution for hadron triggered events, or the distribution of hadrons inside reconstructed jets.

In order to study the effect of imposing a trigger condition on the shower evolution in isolation, we keep the parameters which are subject to other biases, i.e. parton type, initial energy and the strength of the medium modification fixed for the following section, i.e. the test case is always chosen to be a 20 GeV quark, either in vacuum or propagating through a medium such that the line-integrated medium-induced virtuality is $\Delta Q^2 = 5 $ GeV $^2$.

\subsection{Hard track conditions}

\begin{figure}[htb]
\epsfig{file=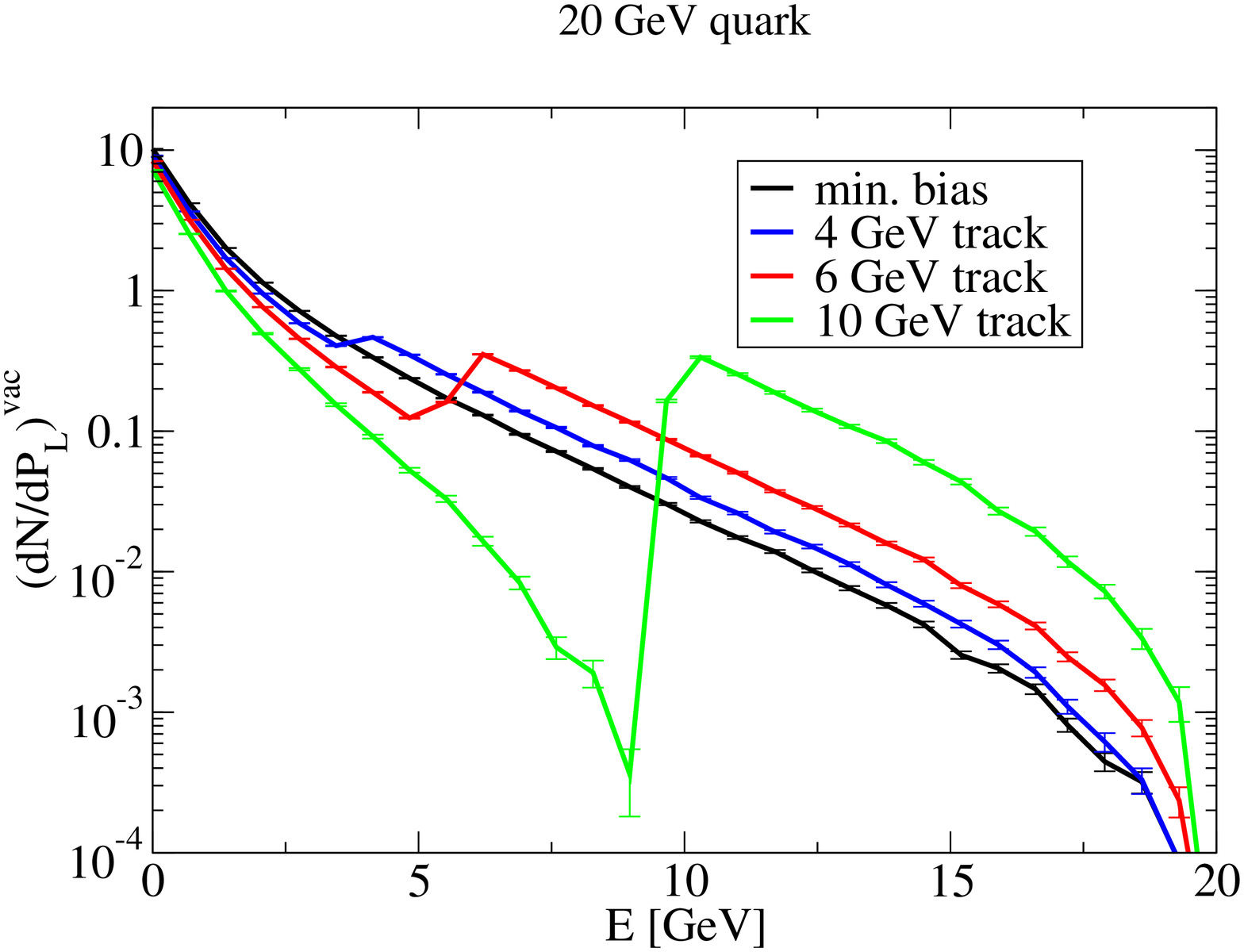, width=8cm}
\caption{\label{F-FF-track}(Color online) Conditional distribution of hadrons at energy $E$ in a shower originating from a 20 GeV quark, given a trigger hadron with the indicated energy in the same shower.}
\end{figure}

Let us now consider showers in which a trigger condition forces a single hard hadron to have the momentum $P_h$. Energy-momentum conservation inside the shower requires to recover (in the medium case approximately) the original shower-initiating parton energy as $E= \sum_i P_i$ (where hadron masses have been neglected), i.e. in the limit where $P_h/E$ is sufficiently large, a significant bias for the remaining distribution is found. This is illustrated in Fig.~\ref{F-FF-track} where the parallel momentum distribution inside a vacuum shower is plotted for various values of the imposed hard track condition.

In essence, a hard track condition leads (by construction) to an enhancement of the hadron yield in the momentum region above the track requirement, and by momentum conservation to a depletion of the yield below. This pattern becomes more pronounced if the trigger hadron takes a sizeable fraction of the total jet energy.

\begin{figure}[htb]
\epsfig{file=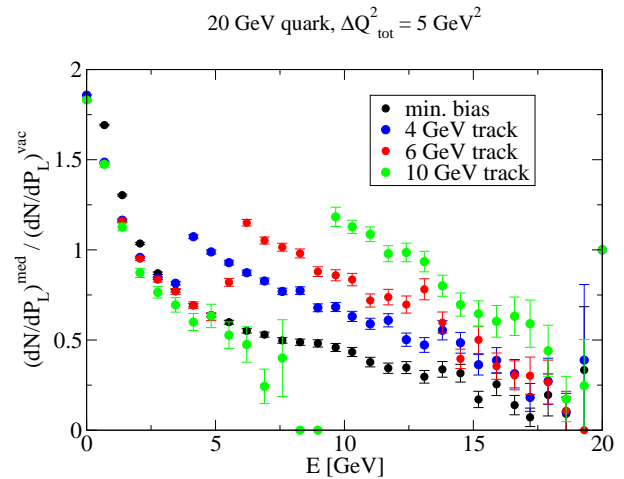, width=8cm}
\caption{\label{F-IAA-track} (Color online) Medium over vacuum ratio of conditional hadron energy distributions in a shower originating from a 20 GeV quark, given a trigger hadron with the indicated energy in the same shower.}
\end{figure}

Fig.~\ref{F-IAA-track} illustrates how the medium modification of the shower responds to a hard track trigger condition at the example of the ratio of the parallel momentum distributions in medium and in vacuum. In the unbiased case, a depletion of the yield at high $P_T$ ('jet quenching') is balanced by a significant yield increase at low $P_T$. A hard track condition tends to remove the depletion at high $P_T$ above the required track momentum. This is a very natural outcome --- while hard tracks are unlikely in the unbiased case and made even less likely by the effect of the medium, a single hard track is always guaranteed once the trigger condition is imposed, and hence it cannot be quenched by the medium. In the presence of such a trigger condition, the medium effect may reduce the rate of triggered events, but it may no longer lead to a quenched high $P_T$ shower pattern. This is a very generic finding --- imposing a trigger bias always tends to reduce the medium modifications of the shower pattern because the trigger condition generates protected structures in the jet.

\subsection{Jet energy conditions}

Another commonly found bias on the shower is the requirement that a jet with at least a certain energy is found. The precise nature of the bias depends on the algorithm used to cluster hadrons into jets and their parameter settings (often an angular radius parameter $R$), as well as on the energy threshold. Qualitatively, it is clear that requiring a substantial flow of the total jet energy into a small cone radius selects showers in which only few branchings take place and as a consequence the parallel spectrum is harder than average while the perpendicular shape is more collimated than average.

\begin{figure}[htb]
\epsfig{file=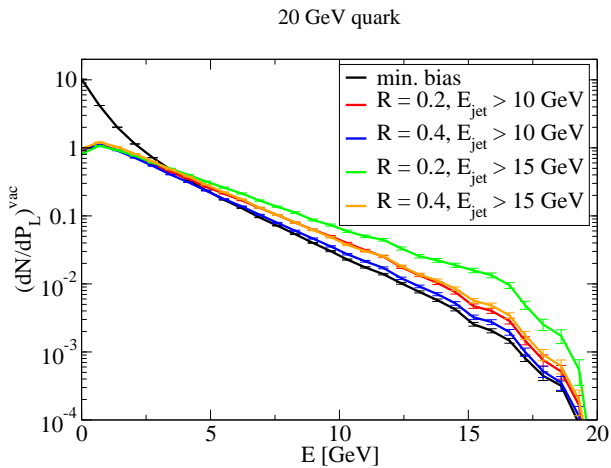, width=8cm}
\caption{\label{F-FF-jet}(Color online) Conditional distribution of hadrons at energy $E$ in a shower originating from a 20 GeV quark, given that the shower after clustering with radius $R$ results in a jet energy above $E_{jet}$.}
\end{figure}

This is shown in Fig.~\ref{F-FF-jet} where the parallel momentum distribution inside a vacuum shower originating from a 20 GeV quark is plotted for various jet energy cuts after the shower has been clustered with anti-$K_T$ for the indicated radius parameter.

As expected, requiring a very collimated jet by asking at least 75\% of the jet energy in a cone of radius $R=0.2$ leads to a sizeable hardening of the hadron spectrum in the jet, with the bias successively decreasing for larger radii or smaller $E_{jet}$. However, unlike in the case of a hard track requirements, there are no pronounced discontinuities in the distribution created by a jet energy condition.

\begin{figure}[htb]
\epsfig{file=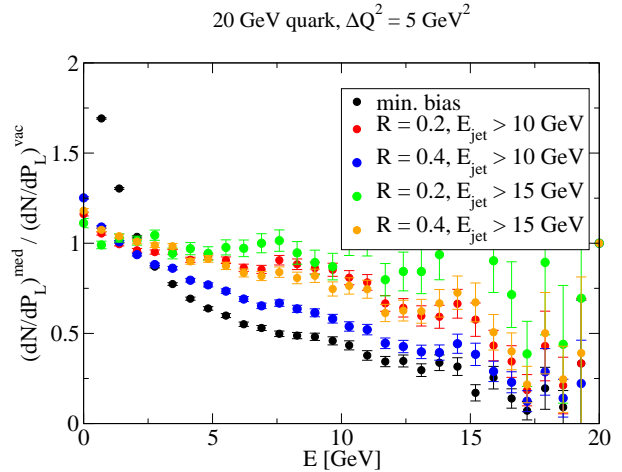, width=8cm}
\caption{\label{F-IAA-jet}(Color online) Medium over vacuum ratio of conditional hadron energy distributions in a shower originating from a 20 GeV quark, given that the shower after clustering with radius $R$ results in a jet energy above $E_{jet}$..}
\end{figure}

Fig.~\ref{F-IAA-jet} illustrates how the medium modification of the shower structure is affected by imposing a jet energy condition. As in the case of a hard track condition, generically a bias on the shower tends to remove the modification, as an increasingly significant part of the shower becomes protected against any modification by the trigger condition.

\subsection{Jet mass conditions}

A final state jet property which is currently not used as a trigger condition in measurements is the mass of a jet. This can be related to the virtuality of the shower-initiating parton which in turn determines the phase space available for vacuum branchings. Thus,  highly virtual partons undergo a much richer branching history before the medium is encountered than hard partons with low initial virtuality. In this way, tagging high jet masses selects events in which configurations of multiple partons undergo medium modification, whereas tagging low mass jets tends to prefer configrations which are dominated by a single leading parton. The strength of medium modifications observed in a shower is thus expected to scale with the jet mass, which can be exploited to get a more differential picture of medium-modified showers \cite{jetMass}.

\section{Case study: away side $I_{AA}$ }

Let us in the following consider a more realistic situation in which the trigger condition refers to a pure final state condition and hence other types of biases (in this section with the exception of the shower bias) occur, in particular kinematic, parton type and geometry bias.

The test case discussed in this section is a measurement of $I_{AA}$ of the conditional yield of hadrons as a function of $P_T$ on the away side (which removes the shower bias), binned as a function of $z_T$ where $z_T = P_T/E_{obs}$. 

Conditional away side yields were first obtained by the STAR collaboration\cite{STAR_dihadron,STAR_dihadron_DzT,STAR_dihadron_PRL} and the strong quenching of the away side correlation peak was almost immediately seen as a spectacular confirmation of the expectations of monojet events in a medium. Similar measurements have now also been performed by the ALICE collaboration at LHC kinematics \cite{ALICE_dihadron} which found somewhat reduced suppression as compared to the RHIC case. While theoretically challenging to compute, dihadron correlations have been a valuable tool to probe for instance pathlength-dependence of energy loss \cite{AdS,ElasticPhenomenology} and to track the fate of subleading hadrons \cite{YaJEM-DE}.

In the case study, $E_{obs}$ is always the trigger energy given the trigger condition. We consider four different trigger trigger conditions: 1) a $\gamma$ ($\gamma$-h) 2) a single hadron (h-h) 3) a jet as defined by STAR  \cite{STAR-jet-h}, including only $\pi^+, \pi^-, \pi^0, K^+, K^-, p, \overline{p}$ or $\gamma$ above 2 GeV clustered with a radius of $R=0.4$ (jet-h) and 4) an ideal jet with all particles clustered into $R=0.4$ (ijet-h).  The trigger energy range is in all cases $12-15$ GeV. 

This selection contains strong kinematical bias (h-h, jet-h) as well as weak kinematical bias (ijet-h, $\gamma$-h), strong parton type bias ($\gamma$-h, h-h) as well as weak parton type bias (jet-h, ijet-h) and strong geometry bias (h-h, jet-h) as well as weak geometry bias ($\gamma$-h, ijet-h).

There is some freedom in the choice of the away side observable, and in principle one could have chosen for instance $P_T$ rather than $z_T$. Each of these choices emphasizes different physics: At low $P_T$, the jet structure is determined largely by the appearance of the medium-induced radiation. As argued in \cite{jet-h}, the enhancement region is essentially set by medium physics and thus is seen at constant $P_T$, not constant $z_T$, in which case plotting the correlation in $P_T$ emphasizes the relevant physics. In contrast, in the high $P_T$ region where $z>0.5$, energy-momentum conservation is a major influence, and the constraints by energy-momentum conservation scale on average approximately with $z_T$ (i.e. as a constant fraction of the trigger energy) rather than $P_T$ (i.e. indepdent of the trigger energy). The choice made here thus emphasizes the high $P_T$ physics at the expense of obscuring the physics of the enhancement due to medium-induced radiation. 
 
Note again that the following results are case studies for the sake of illustration rather than model predictions, since they do not correct for effects like background fluctuations in jet finding and ignore the systematic uncertainty inherent in the choice of the hydrodynamical background, which is known to be important in comparison with real data \cite{HydroSys}.

\subsection{The situation at RHIC}

\begin{figure*}
\epsfig{file=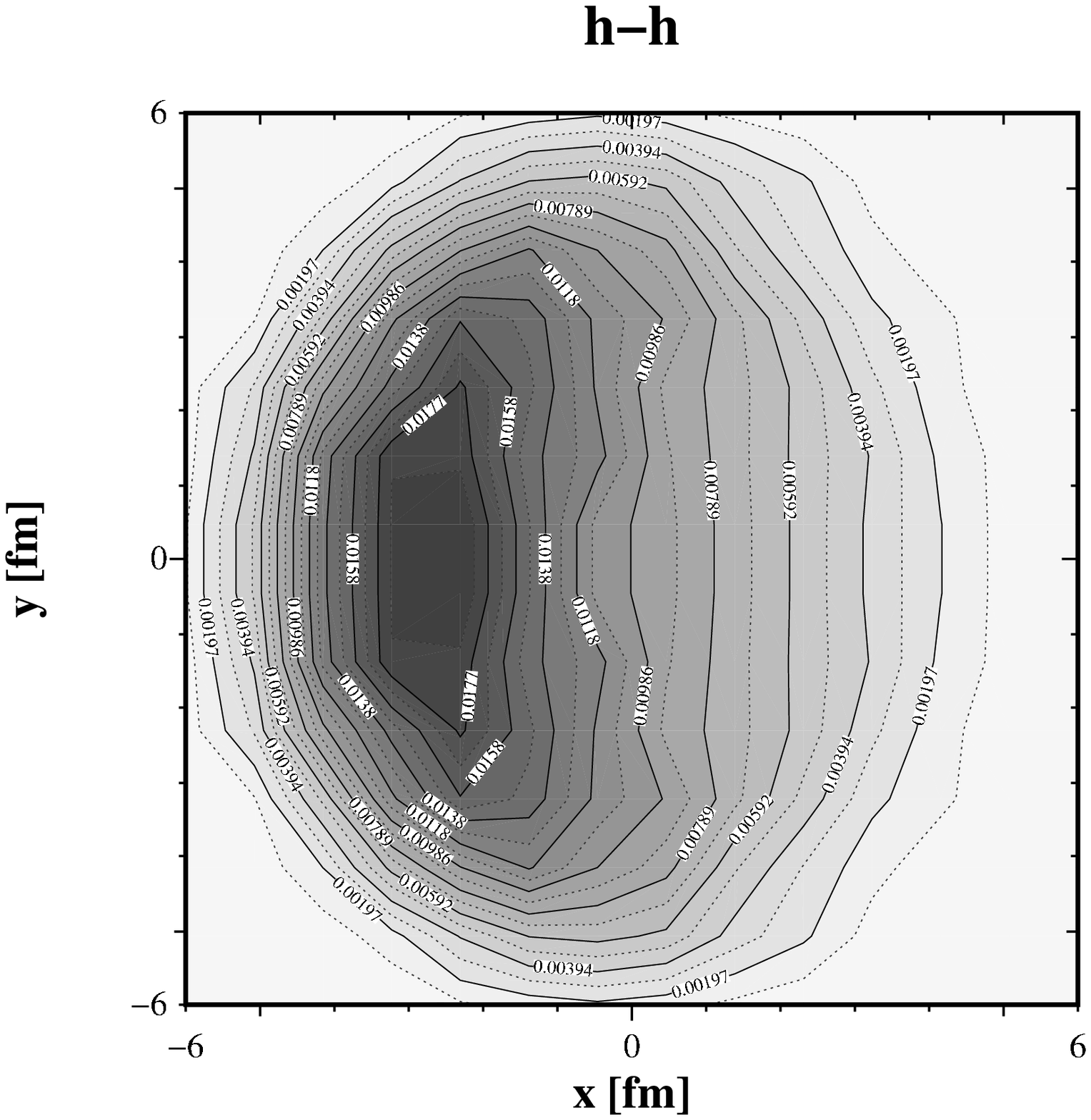, width=5.9cm}\epsfig{file=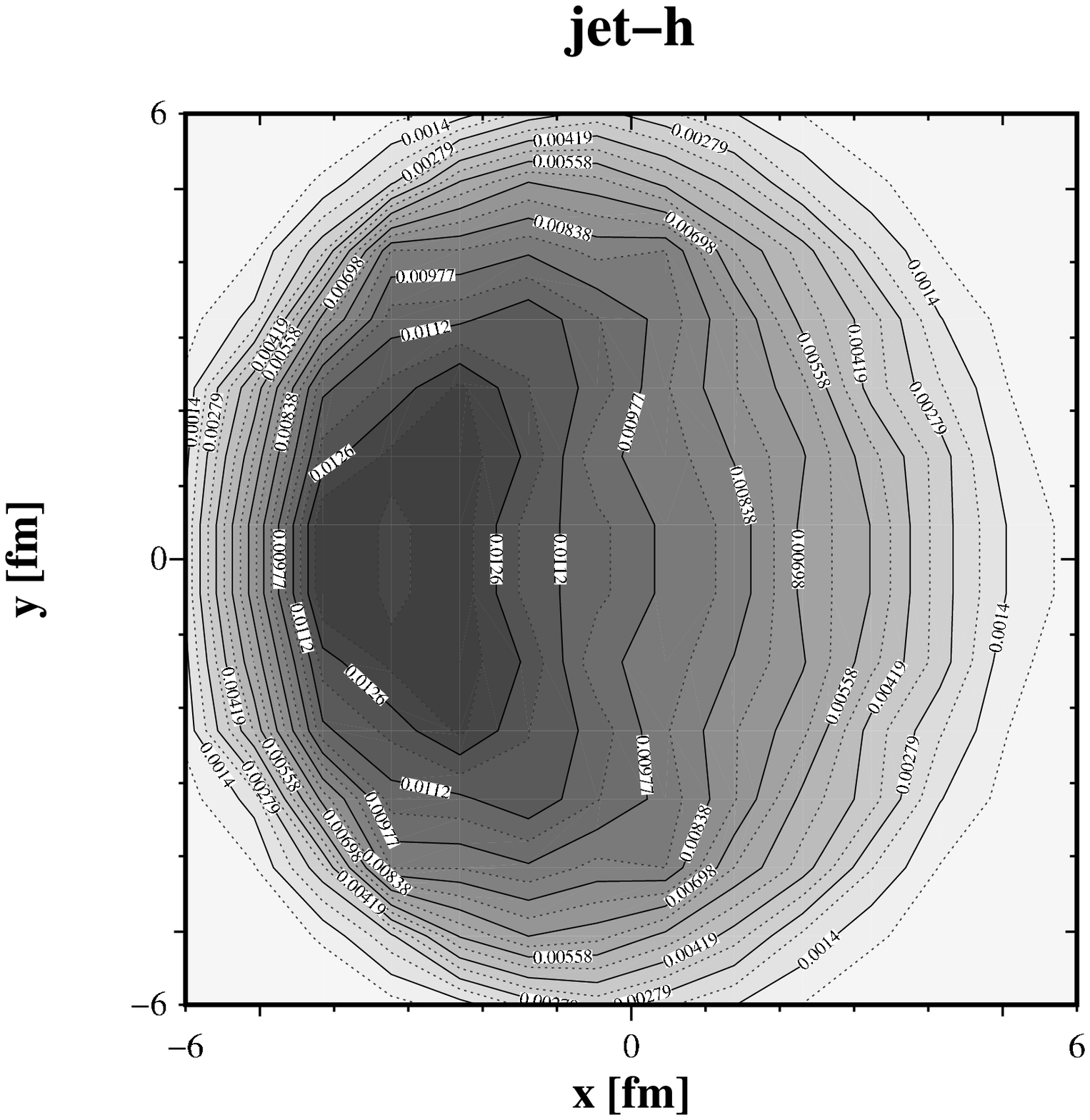, width=5.9cm}\epsfig{file=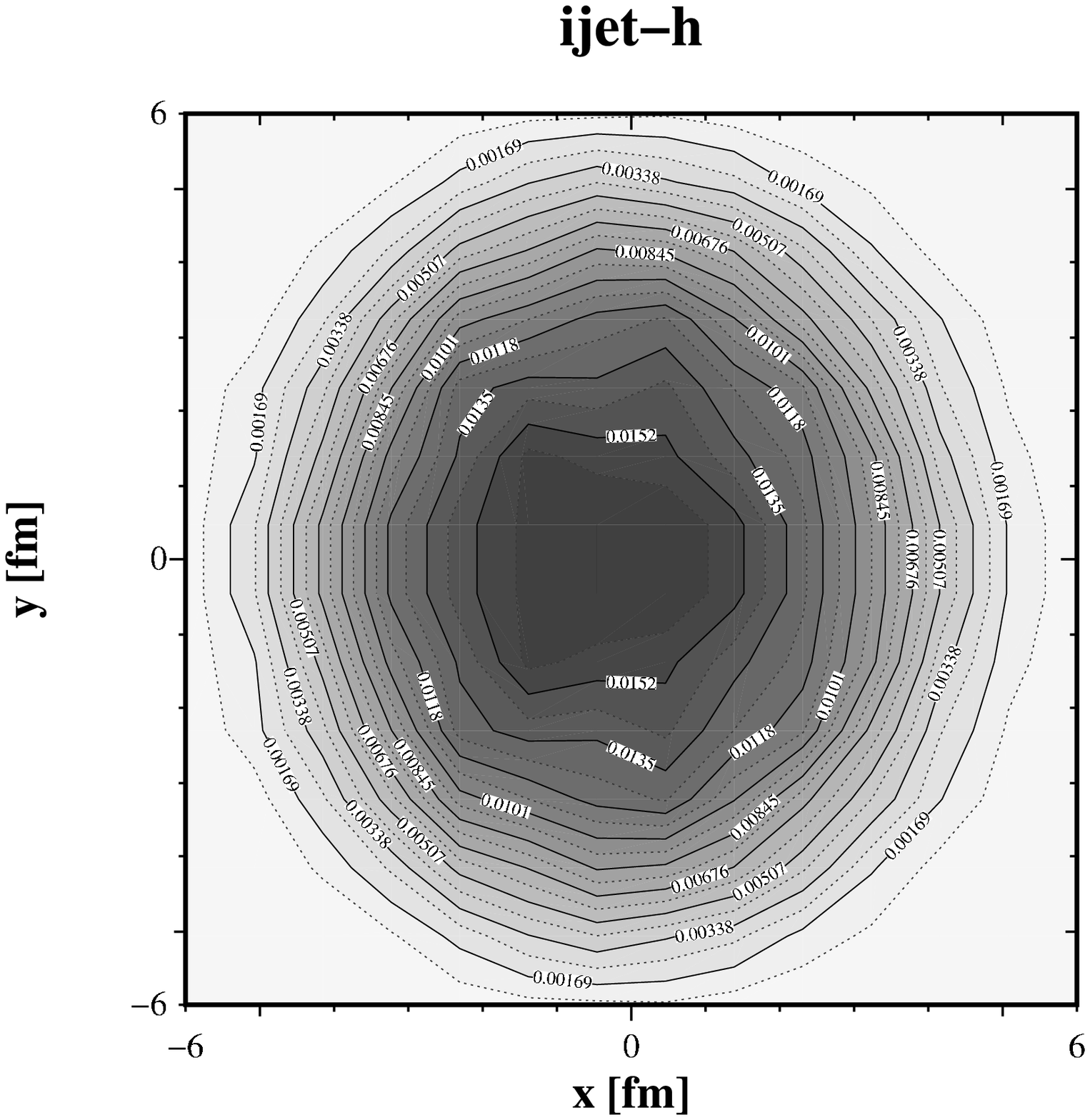, width=5.9cm}
\caption{\label{F-geo-RHIC}Conditional distribution of production vertices in the transverse plane, given a trigger with observed energy $E_{obs}$ between 12 and 15 GeV in 0-10\% central 200 AGeV Au-Au collisions for hadron triggers (left), a jet definition used by STAR (middle) and an idealized jet definition (right). In all cases, the trigger object  momentum vector defines the $-x$ direction.}
\end{figure*}

\begin{figure}[htb]
\epsfig{file=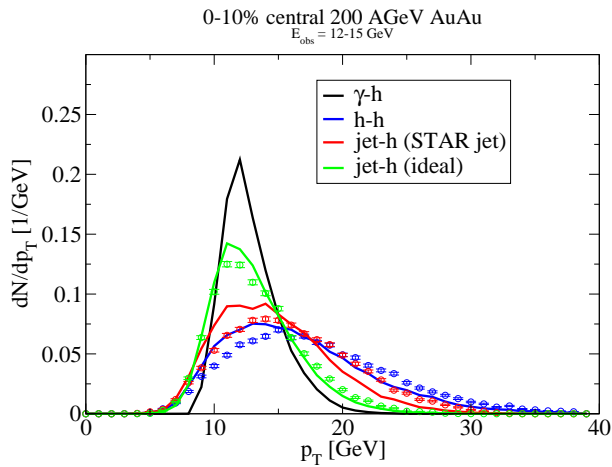, width=8cm}
\caption{\label{F-kinbias-RHIC}(Color online) Conditional momentum distribution of the away side parton given a triggered object in the range of $E_{obs}$ between 12 and 15 GeV for various possibilities for the trigger. Shown for reference is the situation for p-p collisions (lines) as well as the situation in 0-10\% central 200 AGeV AuAu collisions (symbols).}
\end{figure}

We consider first the situation for 0-10\% central Au-Au collisions at RHIC. The distribution of trigger vertices as obtained in the model calculation illustrating the amount of geometrical bias is shown in Fig.~\ref{F-geo-RHIC}, the distribution of away side parton momenta given a trigger in the 12-15 GeV energy range is shown in Fig.~\ref{F-kinbias-RHIC}.

These results confirm in a quantitative way what has been stated earlier: Both h-h and jet-h correlations have a relatively strong geometry bias to trigger on events in which the vertex is close to the surface. This is not so for ijet-h correlations (and since the $\gamma$ does not undergo any final state interaction, the $\gamma$-h correlation has no geometrical bias at all).

At the same time, a $\gamma$-trigger is, up to intrinsic $k_T$ smearing, a relatively faithful representation of the parton kinematics. An ideal jet maps to a somewhat larger region in parton $p_T$, whereas jet-h and h-h probe the widest range in parton kinematics.

\begin{table}[htb]
\begin{tabular}{|l|cccc|}
\hline
trigger & $f_{glue}^{vac}$ near& $f_{glue}^{vac}$ away& $f_{glue}^{med}$ near& $f_{glue}^{med}$ away\\
\hline
$\gamma$-h & N/A & 0.03 & N/A & 0.03\\
h-h & 0.04 & 0.69 & 0.04 & 0.69\\
jet-h & 0.12 & 0.68 & 0.08 & 0.69\\
ijet-h & 0.44 & 0.55 & 0.33 & 0.61\\
\hline
\end{tabular}
\caption{\label{T-gluefrac-RHIC}Conditional fraction of gluon jets on near and away side given a trigger object in the range of $E_{obs}$ between 12 and 15 GeV both in vacuum and in 0-10\% central 200 AGeV Au-Au collisions.}
\end{table}

The parton type bias is summarized in Tab.~\ref{T-gluefrac-RHIC} where the fraction of gluon jets $f_{glue}$ is shown on near and away side in both vacuum and medium. While the away side for the $\gamma$-h trigger is almost a pure quark jet sample, all other trigger conditions lead to a sizeable gluon jet fraction of $\sim 60$\%.

Let us briefly review how these biases affect $I_{AA}$: A strong kinematical bias increases $I_{AA}$ since the available parton energy on the away side increases, giving a larger phase space for particle production. Parton type bias towards gluon jets decreases $I_{AA}$ since gluon jets show softer fragmentation, a comparison of the numbers suggests however that in the particular kinematic window studied here the differences between vacuum and medium are rather small and gluon filtering is not an issue. Finally, a strong geometrical bias decreases $I_{AA}$ since the average in-medium pathlength (and hence the strength of the medium modifications) grows. However, there is no easy {\itshape a priori} argument which would indicate which bias determines the end result.

\begin{figure}[htb]
\epsfig{file=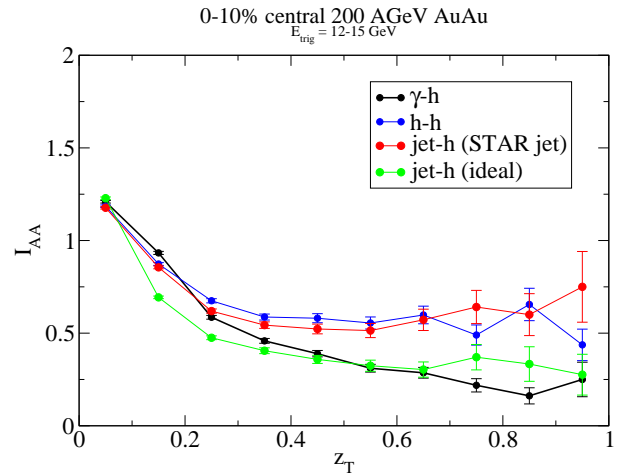, width=8cm}
\caption{\label{F-IAA-RHIC}(Color online) Away side hadron yield modification as a function of $z_T = E_h/E_{obs}$ for various trigger objects in 0-10\% 200 AGeV Au-Au collisions.}
\end{figure}

The actual outcome of the model in terms of away side $I_{AA}$ is shown in Fig.~\ref{F-IAA-RHIC}. Given the fairly sizeable differences in geometry and kinematics probed, the default expectation would be that the resulting $I_{AA}$ exhibits differences to the same degree. However, the actual outcome looks at first glance rather similar. Qualitatively all curves show suppression at high $z_T$ whereas there is enhancement at low $z_T$ (which reflects the generic physics of a MMFF as determined by comparison with a large body of data \cite{Constraining} --- energy lost from hard shower modes is recovered in the enhanced production of subleading hadrons. Quantitatively, there are few differences between $\gamma$-h and ijet-h (which have a markedly different away side population of quark jets). Jet-h is not separable from h-h, in spite of the fact that the underlying kinematics is somewhat different. There is  however a splitting in the high $z_T$ value of $I_{AA}$ between $\gamma$-h and ijet-h on the ond hand and h-h and jet-h on the other hand which reflects the different geometry bias  and/or kinematcial bias. Note however that the split is not very large and in practice might me difficult to resolve within the systematic uncertainties associated with the choice of a hydrodynamical evolution model for the bulk matter.

There are two possible scenarios which can generate the observed similarity between $\gamma$-h and ijet-h: Either a generic effect makes the outcome of the computation insensitive to the details of the bias, or there is an accidential cancellation of biases acting in different directions. 

\begin{figure}[htb]
\epsfig{file=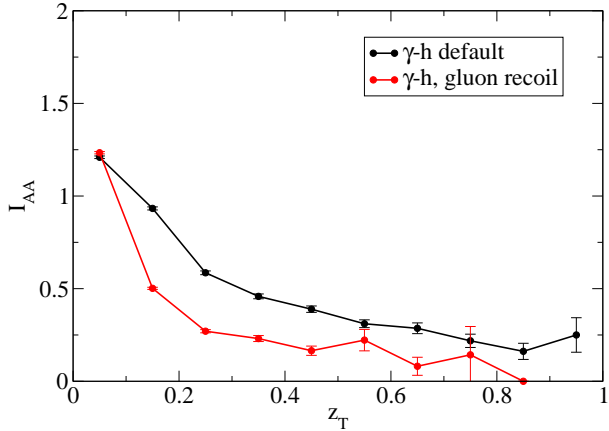, width=8cm}
\caption{\label{F-IAA-gamma-h-RHIC}(Color online)Away side hadron yield modification as a function of $z_T = E_h/E_{obs}$ for a $\gamma$ trigger in 0-10\% 200 AGeV Au-Au collisions, assuming the actual pQCD scattering and a scenario in which only the channel $q\overline{q}\rightarrow g\gamma$ is active.}
\end{figure}

The result shown in Fig.~\ref{F-IAA-gamma-h-RHIC} argues that the latter scenario is true --- if the parton type bias is changed to the (unphysical) case that only gluons recoil from a $\gamma$ trigger, the stronger interaction of the gluon with the medium is expected to lead to additional softening of the away side yield --- which is exactly what is observed. Thus, the observation that $\gamma$-h and ijet-h results fall almost on top of each other is not due to some generic mechanism, but results from a non-trivial cancellation of biases.

This in turn argues that if the relative strength of the biases can be changed experimentally, the cancellation can no longer be expected to occur. One possibility to do so is to consider the LHC kinematic range at a significantly higher $\sqrt{s} = 2.76$ ATeV where we will see differences between $\gamma$-h and ijet-h results (cf. Fig.~\ref{F-IAA-LHC}). 

\subsection{The situation at LHC}

When going from $\sqrt{s} = 200$ AGeV to  $\sqrt{s}$ of 2.76 ATeV with trigger momentum ranges kept fixed, the following trends are expected in the biases: First, the hard collisions probe the nuclear initial state at lower $x \sim 2 E_{obs}/\sqrt{s}$, and consequently there is a transition to a significantly more gluon-dominated regime, as gluons increasingly constitute the largest share of the low $x$ parton distribution. This has an effect on the parton type bias. In addition, the momentum spectrum of produced partons gets much harder, which implies a weakening of the kinematic bias since the 'penalty' for using a very energetic parton to produce a high $P_T$ hadron decrases. As a result, the correlation between parton momentum and jet or leading hadron momentum generically weakens. Finally, there is also a more copious production of bulk matter, both medium temperature and density are increasing with $\sqrt{s}$, which implies a strengthened geometrical bias. However, since the available kinematic range grows $\sim \sqrt{s}$ whereas the medium density grows as a weak power of $\sqrt{s}$ (for instance $\sim \sqrt{s}^{0.574}$ in the EKRT model \cite{EKRT}), there is some reason to expect a net weakening of the geometrical bias.

Again, note that the following results are for illustration and not predictions, as they use a direct extrapolation from RHIC to LHC energies \cite{RAA-LHC} with no attempt to tune model parameters to LHC data or to explore the systematic uncertainty given by the choice of the hydrodynamical background model.

\begin{figure*}
\epsfig{file=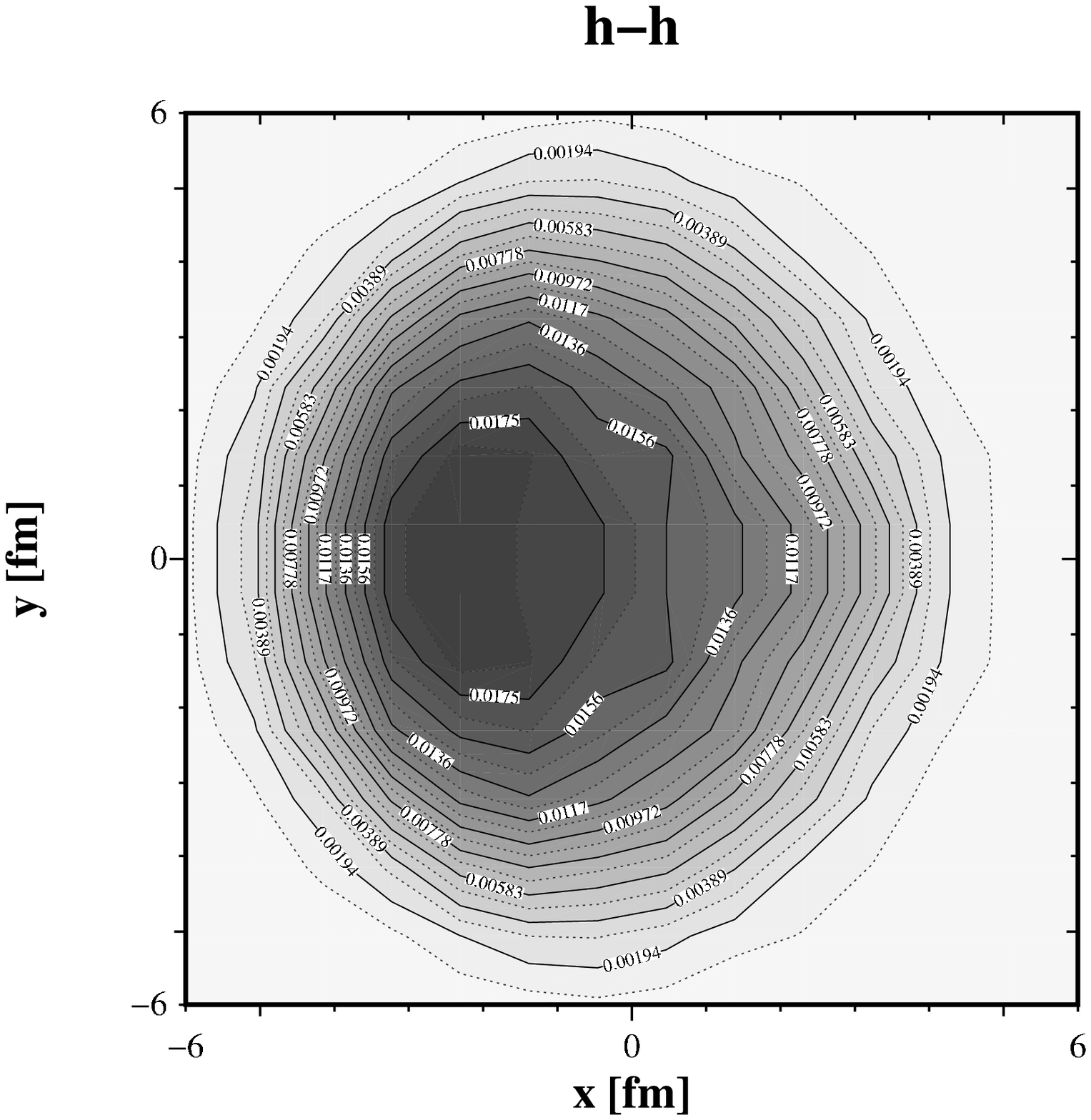, width=5.9cm}\epsfig{file=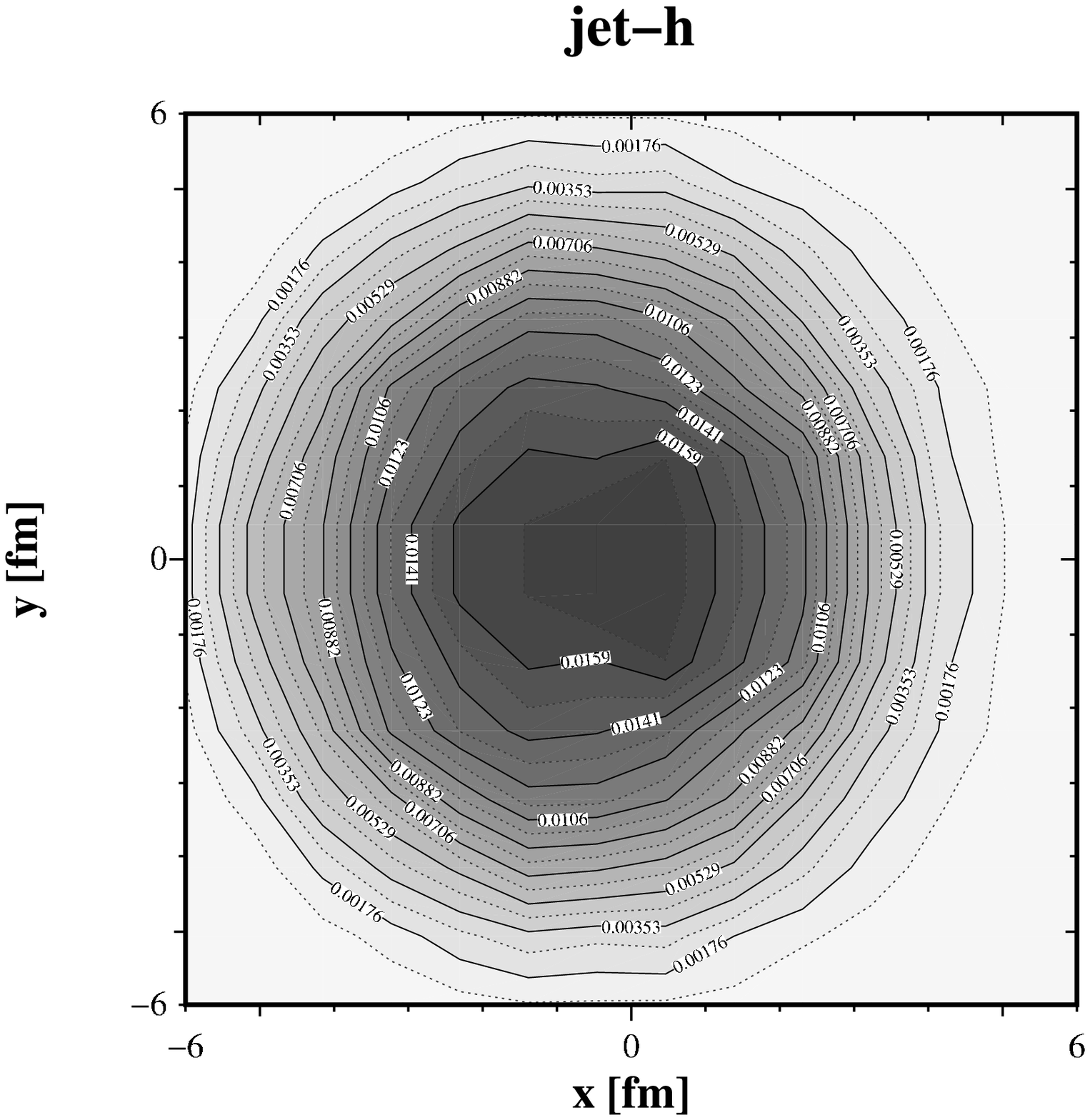, width=5.9cm}\epsfig{file=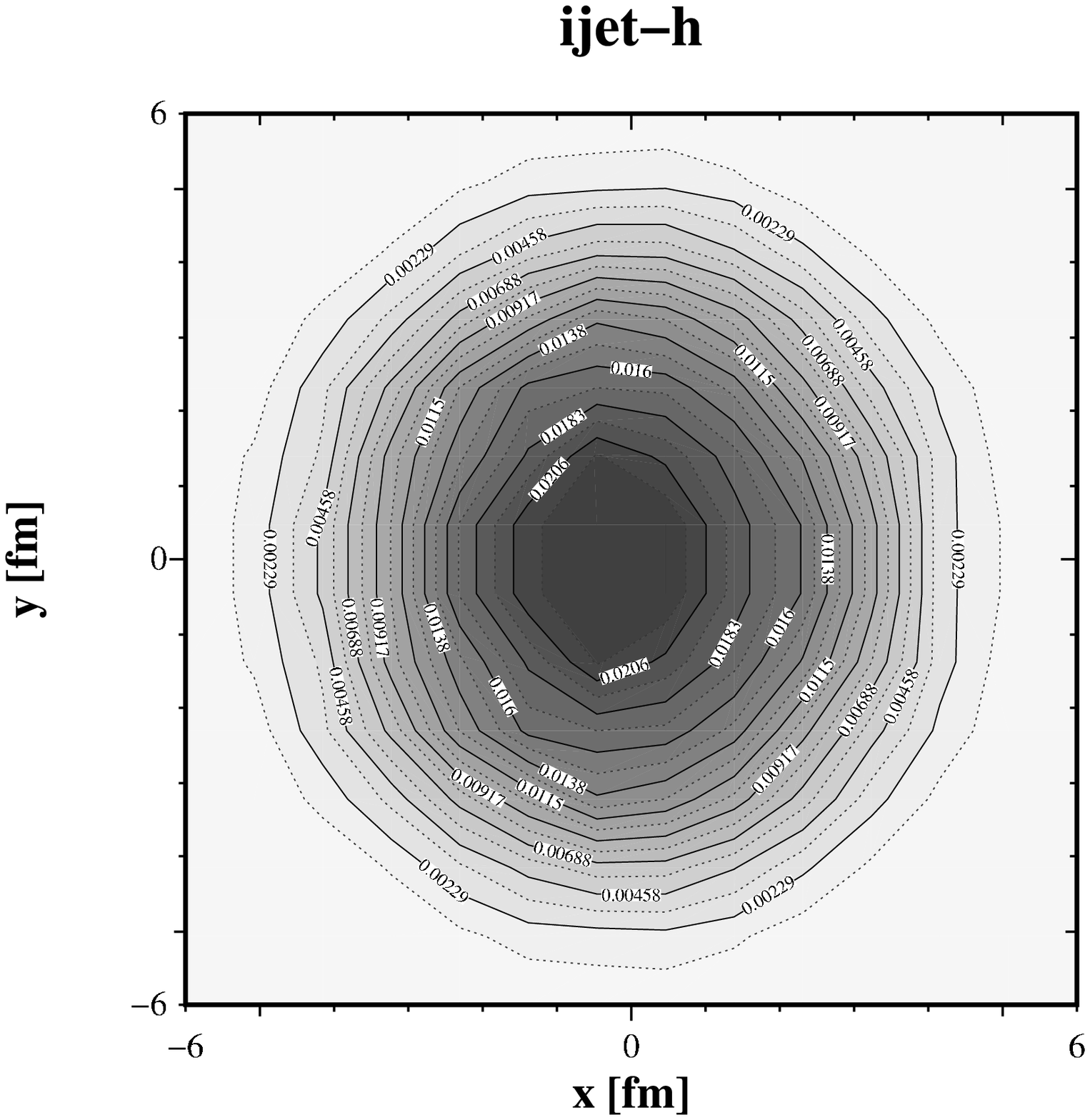, width=5.9cm}
\caption{\label{F-geo-LHC}Conditional distribution of production vertices in the transverse plane, given a trigger with observed energy $E_{obs}$ between 12 and 15 GeV in 0-10\% central 2.76 ATeV Pb-Pb collisions for hadron triggers (left), a jet definition used by STAR (middle) and an idealized jet definition (right). In all cases, the trigger object  momentum vector defines the $-x$ direction.}
\end{figure*}

\begin{figure}[htb]
\epsfig{file=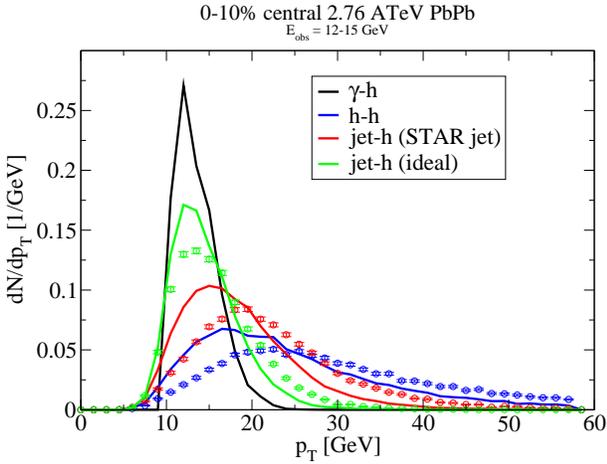, width=8cm}
\caption{\label{F-kinbias-LHC}(Color online) Conditional momentum distribution of the away side parton given a triggered object in the range of $E_{obs}$ between 12 and 15 GeV for various possibilities for the trigger. Shown for reference is the situation for p-p collisions (lines) as well as the situation in 0-10\% central 2.76 ATeV Pb-Pb collisions (symbols). Note the change in the scale of the $x$-axis in comparison with Fig.~\ref{F-kinbias-RHIC}.}
\end{figure}

An explicit computation  of the geometrical bias shown in Fig.~\ref{F-geo-LHC} confirms this expectation --- despite the higher temperature and density of the LHC medium, the resulting bias on geometry is found to be considerably less due to the harder parton spectrum. This can also be seen from Fig.~\ref{F-kinbias-LHC} where the conditional distribution of away side parton momenta given a trigger is shown.

It is evident that the same range in trigger $P_T$ maps to a much wider range in possible parton kinematics at the LHC than at RHIC. The underlying reason is again the reduced penalty for starting with a high parton energy, which in turn is due to the harder primary parton spectrum. The changes in parton type bias are summarized in Table~\ref{T-gluefrac-LHC}.

\begin{table}[htb]
\begin{tabular}{|l|cccc|}
\hline
trigger & $f_{glue}^{vac}$ near& $f_{glue}^{vac}$ away& $f_{glue}^{med}$ near& $f_{glue}^{med}$ away\\
\hline
$\gamma$-h & N/A & 0.04 & N/A & 0.04\\
h-h & 0.33 & 0.79 & 0.32 & 0.78\\
jet-h & 0.47 & 0.79 & 0.38 & 0.80\\
ijet-h & 0.77 & 0.78 & 0.69 & 0.78\\
\hline
\end{tabular}
\caption{\label{T-gluefrac-LHC}Conditional fraction of gluon jets on near and away side given a trigger object in the range of $E_{obs}$ between 12 and 15 GeV both in vacuum and in 0-10\% central 2.76 ATeV Au-Au collisions.}
\end{table}

As expected, it can be seen that in particularly the near side gluon fraction at LHC is much increased over RHIC values, but also that the away side gluon fraction is more independent of the near side gluon fraction. This dependence at RHIC happened because of the dominance of the $gq \rightarrow qg$ reaction, which is no longer dominating at LHC kinematics.

\begin{figure}[htb]
\epsfig{file=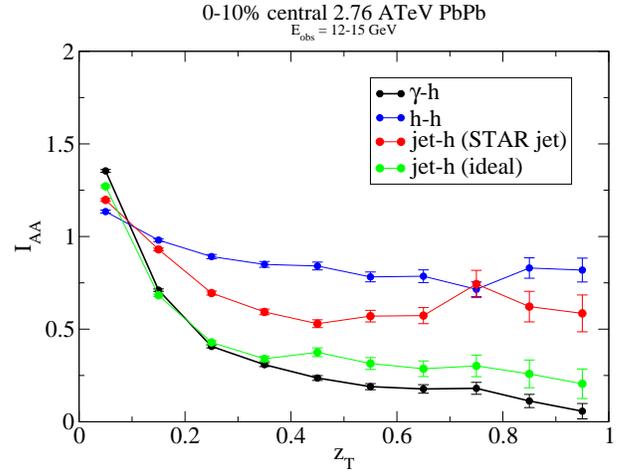, width=8cm}
\caption{\label{F-IAA-LHC}(Color online) Away side hadron yield modification as a function of $z_T = E_h/E_{obs}$ for various trigger objects in 0-10\% 2.76 ATeV Pb-Pb collisions.}
\end{figure}

The final model result in terms of away side $I_{AA}$ are shown in Fig.~\ref{F-IAA-LHC}. These results show that the change in $\sqrt{s}$ from RHIC to LHC, resulting in different kinematical, geometrical and parton type biases is in principle strong enough to leave significant traces in observable quantities. Any similarity between RHIC and LHC results should therefore not be seen as caused by the same generic (and hence trivial) dynamics, but rather as carrying meaningful information in terms of the relative strength of biases and their cancellations.

\section{Case study --- the parallel momentum distribution of jets}

Clustering hadrons into jets has been introduced in the study of hard QCD processes in p-p collisions with the aim of providing an easy comparison between pQCD calculations on the partonic level and the experimentally observed hadronic final state. The basic idea is that clustering largely removes the effect of any soft physics like hadronization or additional soft gluon emission which can not alter the flux of energy and momentum in the shower significantly, and thus a fairly direct comparison of experimental observables  is made possible.

It is doubtful if this still constitutes an advantage in the study of medium-modified showers, since medium modification occurs predominantly driven by the medium temperature scale $T\sim 300-500$ MeV (which is soft), i.e. clustering into jets tends to suppress the very effect one sets out to study. It can be shown that this renders dijet imbalance observations fairly insensitive to even gross features of the parton-medium interaction \cite{myA_J}.

One way to overcome this problem is to analyze the spectrum of particles in the observed jets and hence get a more differential picture. In the language developed above, this corresponds to a situation where in addition to kinematic, parton type and geometry bias also the shower bias is relevant. The test case considered here is an analysis of the parallel momentum distribution of jets in 2.76 ATeV PbPb collisions clustered from hadrons above 1 GeV with anti-$k_T$ using $R=0.3$ with the jet energy $E_{jet}$ required to fall into the range of 100 - 110 GeV (note that this is similar to the fragmentation function analysis by the CMS collaboration \cite{CMS-FF}).

From Fig.~\ref{F-geo-LHC} we may infer that the geometry bias in this situation is weak, and from Fig.~\ref{F-kinbias-LHC} we can see that we may expect partons from the trigger energy threshold $E_{jet}$ to about 1.5-2 times the trigger energy to contribute to the yield of jets in the trigger energy range. The complication due to the shower bias can be estimated from Fig.~\ref{F-IAA-jet}: For parton energies close to $E_{jet}$ (i.e. parton energies around 110 GeV) there is reason to expect a strong bias towards a shower structure which is not medium-modified, for parton energies sufficiently above $E_{jet}$ this bias gradually lessens, with the relative weight of these situations being determined by the combination of kinematical and parton type bias. (Note that Fig.~\ref{F-IAA-jet} is obtained for 20 GeV quarks, however due to the approximally self-similar nature of jets caused by the lack of a scale in the QCD splitting kernels corrections evolve only logarithmically in jet energy).

\begin{figure}[htb]
\epsfig{file=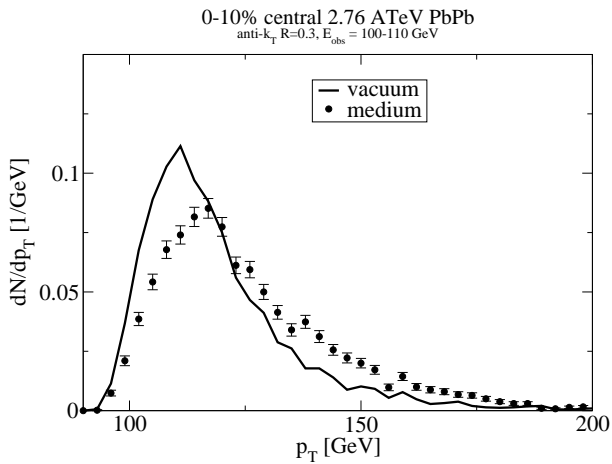, width=8cm}
\caption{\label{F-kinbias-jet-LHC}Conditional momentum distribution of the near side parton given a triggered jet in the range of $E_{obs}$ between 100 and 110 GeV. Shown for reference is the situation for p-p collisions (line) as well as the situation in 0-10\% central 2.76 ATeV Pb-Pb collisions (symbols). }
\end{figure}

The kinematic bias as obtained in the model calculation is illustrated in Fig.~\ref{F-kinbias-jet-LHC}. As expected, parton energies are probed in a range from about 100 to 150 GeV, with a slight shift towards higher energies in the medium case. At the same time, the fraction of gluon jets contributing to the yield in the trigger range decreases from $f_{glue}^{vac} = 0.44$ to $f_{glue}^{med} = 0.36$. A large fraction of jets is hence required to carry 2/3 of the parton energy inside a cone of $R=0.3$, which according to Fig.~\ref{F-IAA-jet} argues for an appreciable shower bias.

\begin{figure}[htb]
\epsfig{file=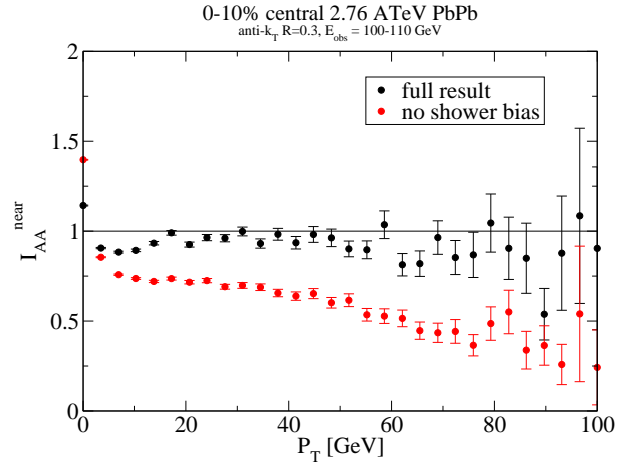, width=8cm}
\caption{\label{F-IAA-jetFF}(Color online) Near side hadron yield medium modification in a $R=0.3$ anti-$k_T$ jet as a function of $P_T$, shown as full result and obtained by neglecting the shower bias.}
\end{figure}

The final result of the model calculation is shown in Fig.~\ref{F-IAA-jetFF} and compared with a computation in which the shower bias effect has been deliberately removed (i.e. the fragmentation is computed for a population of showers as given by the kinematic, geometry and parton type bias as given by evaluating the trigger condition using the full simulation, but it is not checked in this run if a given shower actually clusters to an $E_{jet}$ in the trigger energy range --- note also that without such rejection, the available statistics is much higher).

The results show dramatic differences between taking the shower bias into account or not. In all cases, there is an enhancement of the yield below $P_T \sim3$ GeV (which is not properly resolved by the binning). This is followed by a statistically significant region of depletion in the full calculation which ends at around 30-40 GeV where the full result becomes compatible with unity before statistics runs out. Qualitatively, this agrees with CMS measurements \cite{CMS-FF}. In contrast, the result without shower bias continues to show increasing depletion of the yield up to the highest hadron $P_T$.

The reason for the peculiar pattern of enhancement, depletion and unity observed in the full calculation is a good illustration of the interplay between different biases. At small $P_T$, the contribution of gluon jets is still appreciable, and so the full calculation shows the same enhancement and depletion as the unbiased calculation, albeit driven towards unity by the shower bias. However, at high $P_T$ the yield is almost exclusively due to quark jets since the fragmentation of gluon jets is generically softer. Thus, at some point the enhanced fraction of quark jets in the medium due to gluon filtering leads to a parton type bias towards $I_{AA} > 1$ which happens to approximately compensate the softening of the spectrum due to the medium modification in this kinematic range. The net result is  $I_{AA} \approx 1$ in the high $P_T$ region.

\section{Complicated biases}

In several experimentally relevant situations, even more complicated bias structures appear. One example are 2+1 triggered correlation in which the trigger condition corresponds to observing a coincidence of hard hadrons on both the near and the away side. A different example are triggered or seeded jets in which clustering of the event into jets is only done if a high $P_T$ track has been seen in the event. Let us study these cases in somewhat more detail.

\subsection{2+1 triggered correlations}

\begin{figure*}
\epsfig{file=vdist_h-h_12-15.eps, width=5.9cm}\epsfig{file=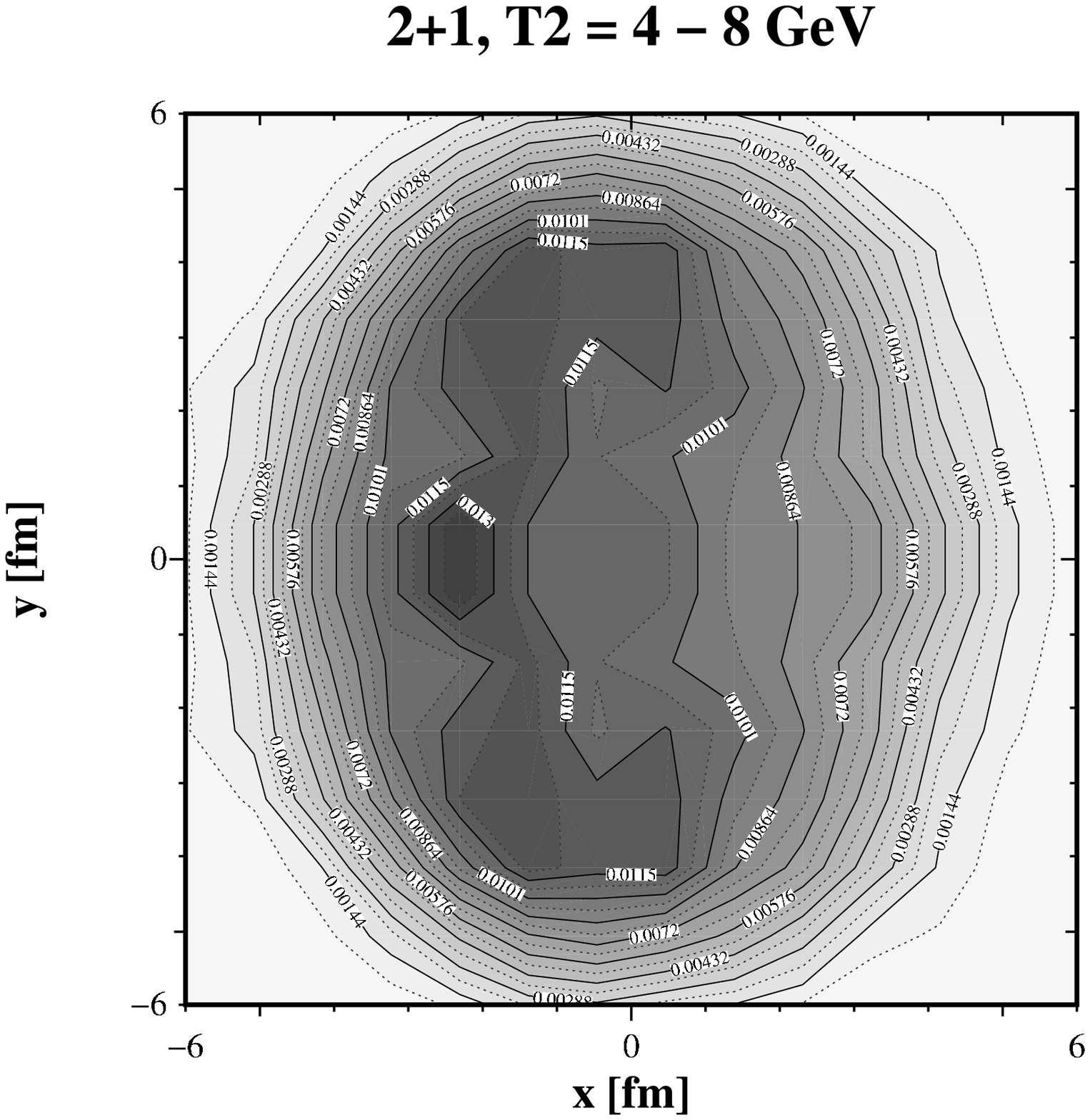, width=5.9cm}\epsfig{file=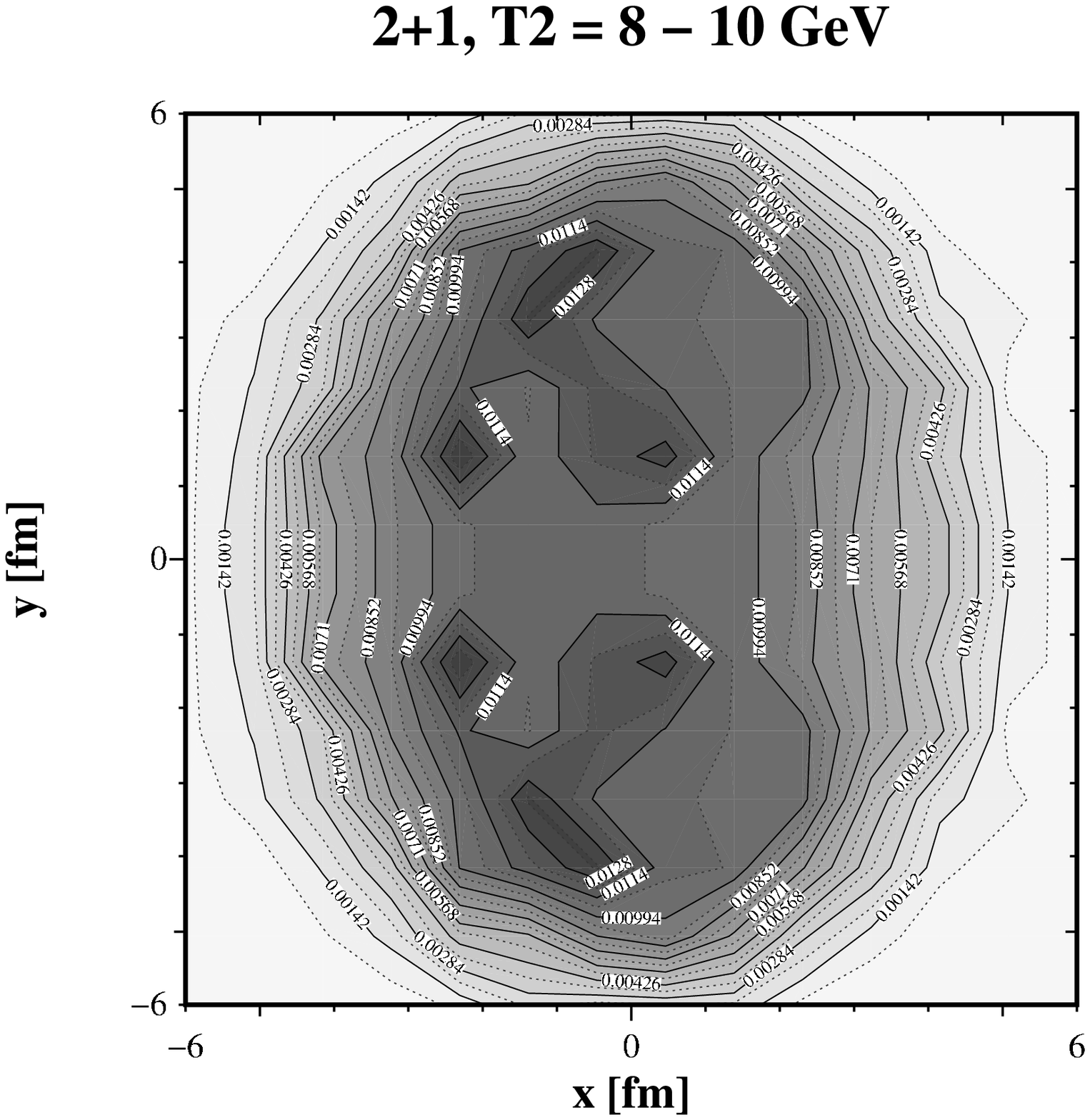, width=5.9cm}
\caption{\label{F-geo-dihadron}Conditional distribution of production vertices in the transverse plane, given a dihadron trigger with observed energy T1 between 12 and 15 GeV in 0-10\% central 200 AGeV Au-Au collisions and T2 set to the indicated values. Shown (left) is also the situation without T2 requirement (see Fig.~\ref{F-geo-RHIC}).  In all cases, the T1  momentum vector defines the $-x$ direction.}
\end{figure*}

While in hadron (or jet) triggered correlations the away side parton propagation is constrained in azimuth to be approximately back to back with the trigger parton, the rapidity of the away side parton is only weakly constrained given the observed rapidity of the near side parton, and only at sufficiently high $P_T$ kinematics forces them to a similar rapidity (see e.g. \cite{MachRap}). The original motivation for introducing 2+1 triggered correlations in which a hard hadron on both the near (T1) and the away side (T2) serves as trigger condition was to explicitly constrain the rapidity of the away side parton, and hence to have a better lever arm to study correlations caused by energy-deposition into the medium. It was however realized fairly quickly that such a trigger condition biases the event towards minimal medium modification of the shower, which tends to make the observation of energy redistribution difficult to impossible \cite{DihadronTrigger}.

Since hard hadron production is rare to begin with, the hard fragmentation in coincidence is an even rarer phenomenon, and this implies a strong bias in the event structure. We may hence expect a strong kinematical bias with on average significantly higher parton energies probed than in the hadron triggered case, a parton type bias leaning towards quark jet coincidences and a symmetric (tangential) geometry bias which minimized the in-medium pathlength for both near and away side parton, combined with a shower bias on each side given by the trigger requirement.

While the original motivation for measuring 2+1 coincidences has a doubtful prospect of being used in practice, 2+1 triggered correlations have the appealing feature that changing the momentum range for $T_2$ allows to change the underlying bias structure in a profound way with minimal effort. The downside is that since hard dihadron coincidences are rare, finite statistics limits their usefulness.

In the following case study, we set T1 to the window of 12-15 GeV in order to compare with previous results for hadron-triggered correlations and study two ranges for T2, 4-8 GeV and 8-10 GeV. We consider the case of 0-10\% central Au-Au collisions at 200 AGeV only.

Fig.~\ref{F-geo-dihadron} shows the geometrical bias obtained from the model calculation. While in the hadron-triggered case there is a surface bias, coming from the requirement of having a short in-medium path for the trigger hadron, with increased momentum required for T2 this gradually changes into a tangential bias for which both near and away side parton in-medium pathlength are minimized. This means that paths through the center of the medium are progressively suppressed and for close to equal momenta of T1 and T2, chiefly the periphery of the medium is probed.

\begin{figure}[htb]
\epsfig{file=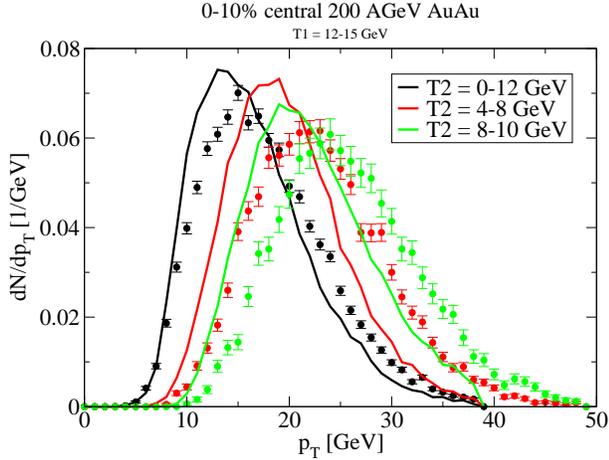, width=8cm}
\caption{\label{F-kinbias-dihadron}(Color online) Conditional momentum distribution of the near side parton given a dihadron trigger with T1 between 12 and 15 GeV and T2 in the indicated range. Shown for reference is the situation for p-p collisions (line) as well as the situation in 0-10\% central 200 AGeV Au-Au collisions (symbols). }
\end{figure}

As expected, the T2 condition has also implications for the parton kinematics. This is demonstrated in Fig.~\ref{F-kinbias-dihadron}. For the highest T2 range, the mean parton momentum probed by the triggered correlation is moved about 10 GeV higher than for the single hadron trigger. The implication of this is naturally a substantial suppression of the trigger rate.

\begin{table}[htb]
\begin{tabular}{|l|cccc|}
\hline
trigger & $f_{glue}^{vac}$ near& $f_{glue}^{vac}$ away& $f_{glue}^{med}$ near& $f_{glue}^{med}$ away\\
\hline
h-h & 0.04 & 0.69 & 0.04 & 0.69\\
T2 = 4-8 GeV & 0.071 & 0.49 & 0.07 & 0.38\\
T2 = 8-10 GeV & 0.10 & 0.29 & 0.05 & 0.20\\
\hline
\end{tabular}
\caption{\label{T-gluefrac-dihadron}Conditional fraction of gluon jets on near and away side given a trigger object in the range of T1 between 12 and 15 GeV and T2 in the indicated range both in vacuum and in 0-10\% central 200 AGeV Au-Au collisions.}
\end{table}

The evolution of the gluon jet fraction with T2 is shown in Tab.~\ref{T-gluefrac-dihadron}. It is apparent that in vacuum the dominance of quark jets leading to a near side with correlated gluons on the away side is progressively broken. This is a natural consequence of the kinematic shift. In the medium, there is a strong gluon filtering effect apparent on the away side, leading to the dominance of correlated quark jets on both near and away side.

\begin{figure}[htb]
\epsfig{file=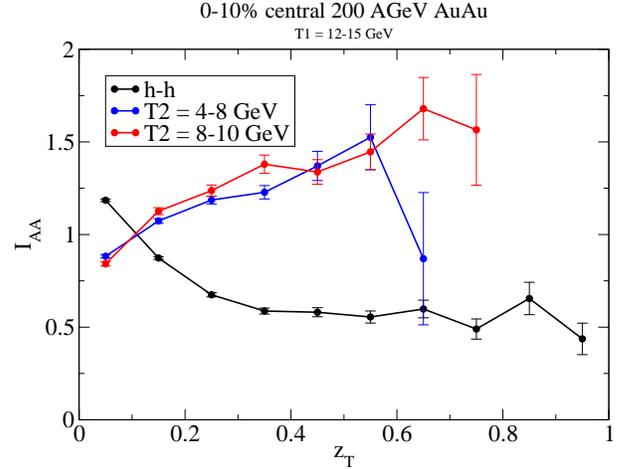, width=8cm}
\caption{\label{F-IAA-dihadrons}(Color online) Away side associated hadron yield medium modification given a single hadron trigger and a dihadron trigger with T1 = 12-15 GeV and T2 in the indicated range in 0-10\% central 200 AGeV Au-Au collisions.}
\end{figure}

The resulting away side $I_{AA}(z_T)$ for the different ranges of T2 is shown in Fig.~\ref{F-IAA-dihadrons}. Perhaps not surprisingly, a significant enhancement of the yield above vacuum is found. This is a result of the strong kinematic bias, shifting parton energy upward and allowing for more phase space for subleading hadron production, the tangential bias which reduces the medium modification for the away side shower as compared with the single hadron triggered case and the parton type bias which drives the away side towards  harder quark jets. The immediate consequence of these biases is a strong reduction in the rate at which triggers are produced (which here reflects in the larger statistical errors for the dihadron triggered results, as significantly more events need to be created for this observable than for hadron triggered events).

\subsection{Jet finding in triggered events}

From an experimental point of view, it is often undesirable to run jet finding algorithms on a set of minimum bias events in heavy-ion collisions, as the vast majority of these events will not contain a hard process and hence significant numerical effort is used to cluster events which are not relevant for the study of hard probes. In such a situation, a triggered event sample where events are only processed further if they contain a hard track or tower (which can be determined early on) can be used. The STAR jet analysis \cite{STAR-jet-h} exemplifies this strategy for instance, whereas jets at ATLAS or CMS do not require such an extra trigger.

However, triggering on events in this way introduces a shower bias. Assuming that in addition it is required that the hard track/tower is part of the leading jet, there is a combined bias from both a jet energy and a track energy condition. Since the kinematical or geometry bias are rather different for jets than for leading hadrons, an interesting question is then whether the objects triggered in this way behave more like jets or like hadrons.

Obviously this depends on the ratio of clustered jet energy to required hadron energy --- if the hadron is in such a momentum regime that a typical jet contains one or more hadrons at this scale, the bias can be expected to be small (for instance, requiring a 5 GeV hadron in a 100 GeV jet is not expected to bias the jet sample in a significant way as the vast majority of 100 GeV jets produces hadrons in the 5 GeV range). On the other hand, once the hadron carries a significant fraction of the total jet energy, jets containing hadrons at such a scale become rare and the additional bias will be substantial.

\begin{figure*}[htb]
\epsfig{file=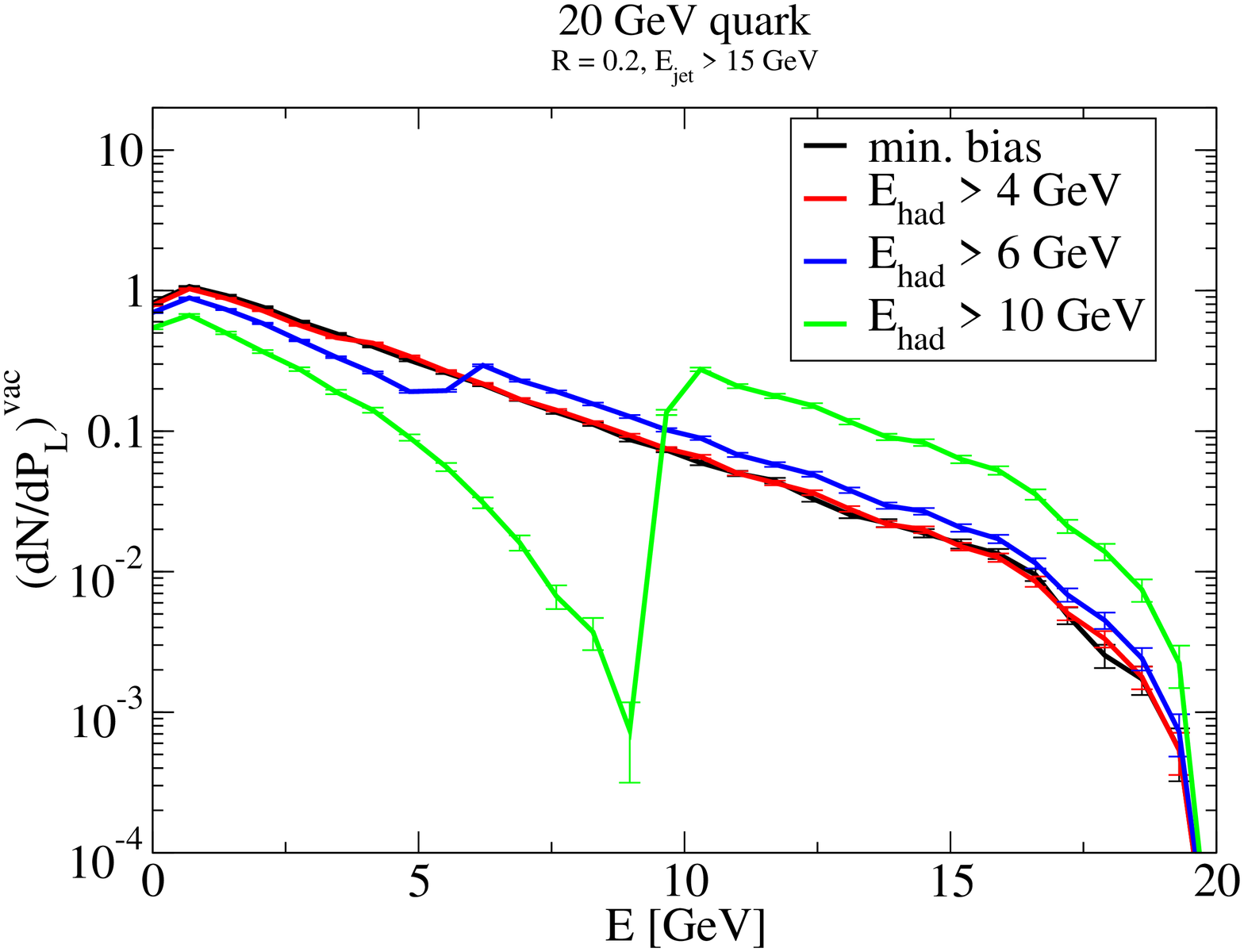, width=8cm}\epsfig{file=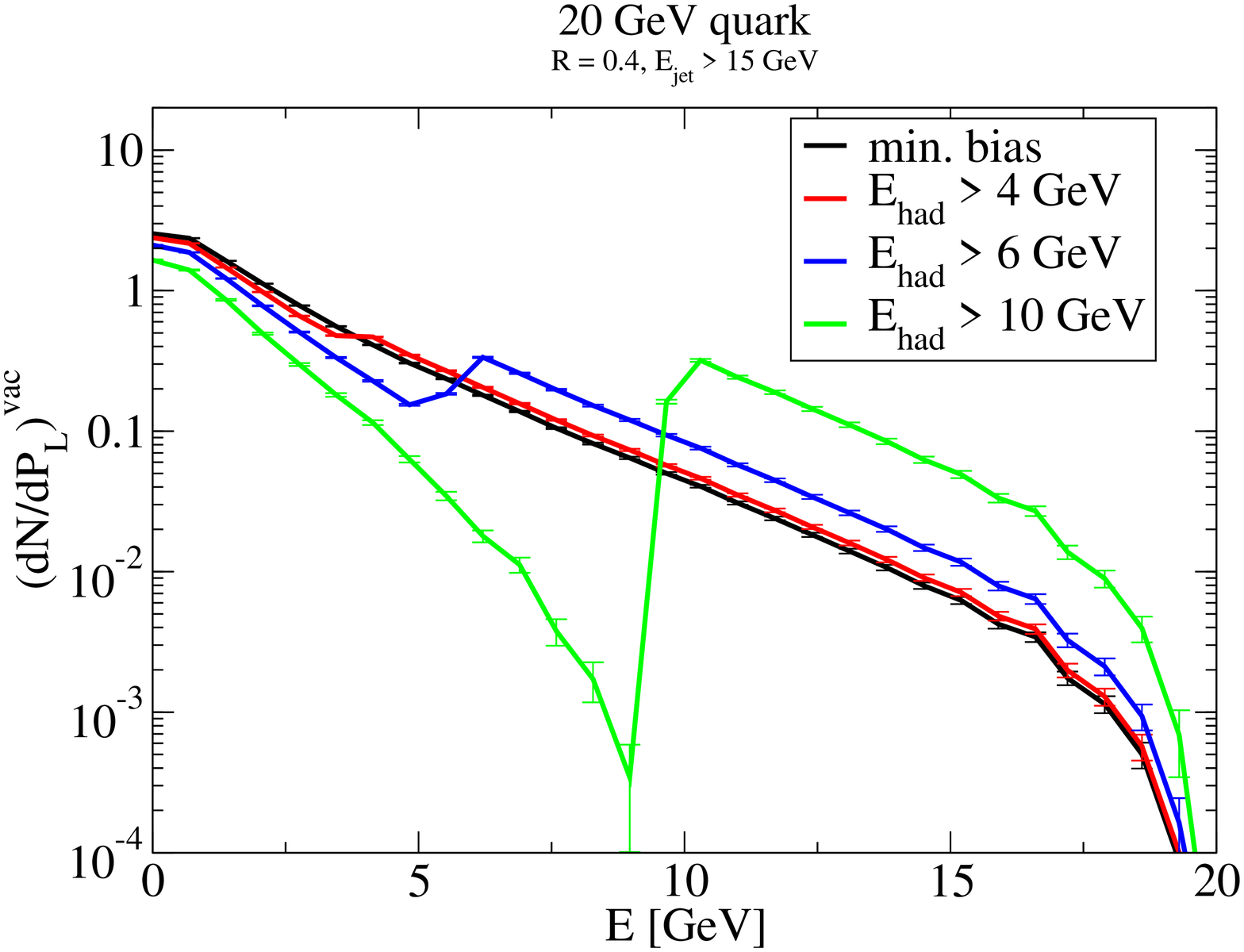, width=8cm}
\caption{\label{F-jet-track}(Color online) Conditional distribution of hadrons at energy $E$ in a vacuum shower originating from a 20 GeV quark, given that the shower clustered using anti-$k_T$ to an energy $E_{jet}>15$ GeV and a trigger hadron with the indicated energy in the same shower for $R=0.2$ (left) and $R=0.4$ (right).}
\end{figure*}

In Fig.~\ref{F-jet-track} sample calculations with a combined shower bias are shown. The rule of thumb emerging from this and similar studies appears to be that the additional bias becomes relevant once the hadron energy reaches about half of the jet energy, with a fairly weak dependence on the radius used to cluster the jet. Here it has been tacitly assumed that kinematics is such that jet finding typically recovers ~75\% of the parton energy, which according to Fig.~\ref{F-Pz} is not a bad assumption.

\section{Conclusions}

\subsection{Biases are everywhere}

As the results of this work show, biases occur for almost any high $P_T$ observable that is in any sense related to a conditional probability --- be it that an explicit trigger condition is evaluated for the event or be it an implicit condition that a jet needs to be clustered before it can be analyzed. This means that understanding and discussing biases is an integral part of any theoretical analysys of hard probes.

The main structure of the biases involved is usually already apparent in the vacuum, and the medium modification to the bias structure of the problem can in many cases be regarded as a correction. The strength of the medium-induced bias is always apparent from the modification (in most cases suppression) of the trigger rate, which in turn is directly measured in disappearance observables such as nuclear modification factors $R_{AA}$ for various trigger objects.

However, the strength of the medium-induced bias does not provide an {\itshape a priori} indication of the modification of  conditional yield observables --- some biases (for instance the kinematic bias) may lead to {\itshape increasing} conditional yields despite a suppression of the trigger rate, whereas other biases such as the geometry bias work towards a suppression of conditional yields.

\subsection{Biases are important}

As for instance Fig.~\ref{F-IAA-jetFF} indicates, taking biases into account properly can change the result of a computation quantitatively and even qualitatively. Thus, a theoretical model calculation in which the fate of unbiased parton showers in a medium is obtained can not be expected to compare with data based on the notion that the bias of finding a jet is somehow small or would not influence the results in a relevant way. In the particular case of the parallel intra-jet hadron distribution which experimentally appears unchanged in a medium over a large momentum range \cite{CMS-FF}, the naive comparison with theory without shower bias would find a large discrepancy to the data, and hence lead to the conclusion that a previously not considered mechanism which makes shower evolution in medium similar to vacuum needs to be introduced. However, taking the shower bias into account properly, the need for any additional mechanism goes away.

Biases are at least equally important for triggered correlation measurements --- however for these this is usually expected, although the relative strength of different biases can lead to counter-intuitive results when one e.g. expects suppression of a yield based on the geometry bias whereas in the actual situation the kinematical bias dominates, leading to a net enhancement.

\subsection{Use of biases}

Biases can appear as a nuisance in cases where they suppress the physics one is interested in studying, perhaps the most striking illustration is the shower bias for a hadron or jet trigger (see Figs.~\ref{F-IAA-track} and \ref{F-IAA-jet}) where strong medium modifications which are {\itshape a priori} present in the shower are suppressed by the bias with the effect that $I_{AA}$ is driven towards unity. Such nuisance biases should be avioded if possible --- in the context of shower biases, this can be done at the simple expense of separating trigger side and observable side, i.e. study away side jets with a trigger hadron on the near side (as suggested e.g. in \cite{Peter_Jets}).

However, in many cases biases can be utilized by the design of a measurement to control the relevant parameters of the hard process to specifically probe the dependence of a medium-modification on a single control parameter. As an example, consider for instance a comparison of jet-h and $\gamma$-h correlations at RHIC kinematics. According to Fig.~\ref{F-geo-RHIC}, the geometrical bias of a sufficiently inclusive jet definition is very weak, i.e. in this case $\gamma$-h and ijet-h correlations probe almost the same geometry. According to Fig.~\ref{F-kinbias-RHIC}, they also have a fairly similar kinematical bias, and a small shift in trigger energy range can make the underlying parton enery on average the same. The main difference between the two situations is then given by Tab.~\ref{T-gluefrac-RHIC} from which one can read off that the $\gamma$-h correlation produces a high fraction of quark jets on the away side whereas the ijet-h correlation is dominated by away side gluon jets. Thus, in a measurement the bias can be designed to specifically probe the different evolution of quark and gluon jets in the medium.

In a similar way, the geometry bias can be systematically varied by changing the constituent cut used for the clustering (cf. Fig~\ref{F-geo-RHIC}). If the variation of parton kinematics associated with the change is compensated for by a change in the trigger energy range, the measurement can be made to probe various regions of the medium selectively.

Of course, such designed observables are inevitably to some degree model-dependent. However, in most cases the dominant structure of e.g. kinematical or parton type bias is given by vacuum QCD, on top of which the medium-induced bias is a correction. This means that parameters like the necessary shift in trigger energy range can be determined approximately by well-known vacuum physics and do not have to rely on a particular model of parton-medium interaction in a significant way.

\section{Summary}

The ancient Chinese strategist Sun Tzu writes in his 'Art of War': 

{\itshape It is said that if you know your enemies and know yourself, you will not be imperiled in a hundred battles; if you do not know your enemies but do know yourself, you will win one and lose one; if you do not know your enemies nor yourself, you will be imperiled in every single battle.} 

Similarly, one might summarize the results of this work as:

 {\itshape If you understand parton medium interaction and the involved biases, everything will become clear. If you understand parton-medium interaction but not the biases, some observables will make sense, others will appear as puzzles; but if you have neither an understanding of biases nor a good model of parton-medium interaction, you can not know the implication of any hard probe.}

\begin{acknowledgments}
 
I'd like to thank H.~Caines, J.~Putschke, P.~Jacobs, A.~Ohlson, M.~Connors, O.~Evdokimov, B.~Cole, G.~Roland, P.~Steinberg, M.~van Leeuwen and K.~Loizides for interesting discussions which all in some way led to this paper. This work is supported by the Academy researcher program of the
Academy of Finland, Project No. 130472. 
 
\end{acknowledgments}

\end{document}